%% file: paper.tex
\documentclass[sigconf]{acmart}
\usepackage{comment}
\usepackage{xspace}
\usepackage{listings}
\usepackage{color}
\usepackage[ruled, vlined]{algorithm2e}
\usepackage{float}
\usepackage{pgfplots}
\usetikzlibrary{matrix}
\usetikzlibrary{patterns}
\usepgfplotslibrary{groupplots}
\pgfplotsset{compat=newest}
\usepackage{xcolor}
\usepackage{soul,color}
\usepackage{multirow}
\usepackage{adjustbox}
\usepackage{array}
\usepackage{placeins}

\soulregister\ref7
\soulregister\cite7

\DeclareMathDelimiter{(}{\mathopen} {operators}{"28}{largesymbols}{"00}
\DeclareMathDelimiter{)}{\mathclose}{operators}{"29}{largesymbols}{"01}

\SetAlFnt{\footnotesize}
\SetCommentSty{mycommfont}
\DontPrintSemicolon

\definecolor{lightgray}{rgb}{.95,.95,.95}
\definecolor{darkgray}{rgb}{.4,.4,.4}
\definecolor{purple}{rgb}{0.65, 0.12, 0.82}
\definecolor{mygreen}{rgb}{0,0.6,0}
\lstdefinelanguage{Assembly}{
    keywords={call, push, struct, and, brk, read, write, cmp, mov, add, ret, false, for, if, else, in, int, lea, then, continue, dec, jnz, while, char, printf, void},
    keywordstyle=\color{blue}\bfseries,
    ndkeywords={assert, condition, class, export, boolean, throw, implements, import, this},
    ndkeywordstyle=\color{red}\bfseries,
    identifierstyle=\color{black},
    sensitive=false,
    comment=[l]{//},
    morecomment=[s]{/*}{*/},
    commentstyle=\color{purple}\ttfamily,
    stringstyle=\color{red}\ttfamily,
    morestring=[b]',
    morestring=[b]"
}
\definecolor{mGreen}{rgb}{0,0.6,0}

\lstdefinestyle{CStyle}{
	keywordstyle=\color{blue},
	backgroundcolor=\color{white},
	commentstyle=\color{mGreen},
	extendedchars=true,
	basicstyle=\footnotesize\ttfamily,
	showstringspaces=false,
	showspaces=false,
	numbers=left,
	numberstyle=\footnotesize,
	numbersep=3pt,
	tabsize=2,
	breaklines=true,
	showtabs=false,
	captionpos=b,
	xleftmargin=1em,
	frame=single,
	framexleftmargin=1.5em,
	escapeinside={@}{@},
	language=C
}

\usepackage{fancyvrb}

\DefineVerbatimEnvironment{pcode}{Verbatim}{
	numbers=left,
	xleftmargin=5mm
}

\newcommand{\codename}{\fazer}
\newcommand{\fazer}{\textsc{Sivo}\xspace}
\newcommand{\go}{GreyOne\xspace}
\newcommand{\taint}{\texttt{TaintFAST}\xspace}

\newcommand{\ivica}[1]{{\textcolor{blue}{(Ivica: {#1})}}}

\renewcommand{\paragraph}{\vspace{2pt}\noindent \textbf}

\setcopyright{none} 
\acmConference[Anonymous Submission to ACM CCS 2021]{ACM Conference on Computer and Communications Security}
\acmYear{2019}

\begin{document}

\date{}

\title{Refined Grey-Box Fuzzing with \fazer}

\input{chapters/abstract}

\author{Ivica Nikoli\'c}
\affiliation{
  \institution{School of Computing, NUS}            
  \country{Singapore}                    
}

\author{Radu Mantu}
\affiliation{
  \institution{University Politehnica of Bucharest}            
  \country{Romania}                    
}

\author{Shiqi Shen}
\affiliation{
  \institution{School of Computing, NUS}            
  \country{Singapore}                    
}

\author{Prateek Saxena}
\affiliation{
  \institution{School of Computing, NUS}            
  \country{Singapore}                    
}


\maketitle


\input{paper-body} 


\bibliographystyle{ACM-Reference-Format}
\bibliography{paper}

\begin{appendix}

\input{paper-appendix} 

\end{appendix}

\end{document}

%% file: chapters/abstract.tex

\begin{abstract}

We design and implement from scratch a new fuzzer called \fazer that refines multiple stages of grey-box fuzzing. 
First, \fazer refines data-flow fuzzing in two ways: (a) it provides a new taint inference engine that requires only logarithmic in the input size number of tests to infer the dependency of all program branches on the input bytes, and (b) it deploys a novel method for inverting branches by solving directly and efficiently systems of inequalities. 
Second, our fuzzer refines accurate tracking and detection of code coverage with simple and easily implementable methods. 
Finally, \fazer refines selection of parameters and strategies by parameterizing all stages of fuzzing and then dynamically selecting optimal values during fuzzing. Thus the fuzzer can easily adapt to a target program and rapidly increase coverage.  
We compare our fuzzer  to $11$ other state-of-the-art grey-box fuzzers on $27$ popular benchmarks. Our evaluation shows that \fazer scores the highest both in terms of code coverage and in terms of number of found vulnerabilities.  

\end{abstract}

%% file: paper-body.tex
\input{chapters/intro}
\input{chapters/problem2}

\input{chapters/refinements}

\input{chapters/design}

\input{chapters/implementation}

\input{chapters/refine-limitations}
\input{chapters/eval}

\input{chapters/related-work}

\input{chapters/conclusion}
\input{chapters/ack}

%% file: chapters/intro.tex

\section{Introduction}
\label{sec:introduction}

Fuzzing is the automatic generation of test inputs for programs with the goal of finding bugs. With increasing investment of computational resources for fuzzing, tens of thousands of bugs are found in software each year today. We view fuzzing as the problem of maximizing coverage within a given computational budget. The coverage of all modern fuzzers improves with the computation budget allocated. Therefore, we can characterize the quality of a fuzzer based on its {\em rate of new coverage}, say, the number of new control-flow edges exercised, per CPU cycle on average.  


Broadly, there are three types of fuzzers. Black-box fuzzers do not utilize any knowledge of the
program internals, and are sometimes referred to as undirected fuzzers. White-box fuzzers perform intensive instrumentation, for example, 
enabling dynamic symbolic execution to systematically control which program branches to invert in each test. Grey-box fuzzers introduce low-overhead instrumentation into the tested program to 
guide the search for bug-triggering inputs. These three types of fuzzers can be combined. For instance, recent hybrid fuzzers selectively utilize white-box fuzzers in parallel to stand-alone grey-box fuzzers. 
Of the three types of fuzzers, grey-box fuzzers have empirically shown promising cost-to-bug ratios, thanks to their low overhead techniques, and have seen a flurry of improved strategies.
For example, recent grey-box fuzzers have introduced many new strategies to prioritize seed selection, byte mutations, and so on during fuzzing. Each of these
strategies work well for certain target programs, while being relatively ineffective on others. There is no dominant strategy that works better than all others on all programs presently.

In this paper, we present the design of a new grey-box fuzzer called \codename that {\em generalizes} well across many target programs. \codename
embraces the idea that there is no one-size-fits-all strategy that works universally well for all programs. Central to its design is a "parameterization-and-optimization" 
engine to which many specialized strategies and their optimization parameters can be specified. The engine dynamically selects between the specified 
strategies and optimizes their parameters on-the-fly for the given target program based on the observed coverage. The idea of treating fuzzing as an optimization problem
is not new---in fact, many prior fuzzers employ optimization either implicitly or explicitly, but they do so {\em partially}~\cite{lyu2019mopt,yue2020ecofuzz,bohme2017coverage,she2020mtfuzz}. \codename differs from
these works conceptually in that it treats parameterization as a first-class design principle---all of its internal strategies are parameterized. The selection of strategies and determination of
all parameter values is done dynamically. We empirically show the power of embracing {\em complete parameterization} as a design principle in grey-box fuzzers.

\codename introduces $3$ additional novel refinements for grey-box fuzzers.
First, \codename embodies a faster approximate taint inference engine which computes
taint (or sensitivity to inputs) for program branches during fuzzing, using number of
tests that are only logarithmic in the input size. Such taint information is helpful
for directed exploration in the program path space, since inputs influencing
certain branches can be prioritized for mutation. Our proposed refinement improves exponentially over a recent procedure to calculate taint (or data-flow dependencies) during fuzzing~\cite{gan2020greyone}. 
Second, \codename introduces a light-weight form of symbolic interval reasoning which, unlike full-blown
symbolic execution, does not invoke any SMT / SAT solvers. Lastly, it eliminates deficiencies in the calculation
of edge coverage statistics used by common fuzzers (e.g. AFL~\cite{afl}), thereby allowing the optimization procedure to be more effective.
We show that each of these refinements improves the rate of coverage, both individually and collectively.

We evaluate \codename on $27$ diverse real-world benchmarks comprising 
several used in recent work on fuzzing and in Google OSS-fuzz~\cite{ossfuzz}.
We compare \codename to $11$ other state-of-the-art grey-box
fuzzers. We find that \codename outperforms all fuzzers in terms of coverage on $25$ out of the $27$ benchmarks we tested. 
Our fuzzer provides $20\%$ increase in coverage compared to the next best fuzzer, and $180\%$ increase compared to the baseline AFL. Furthermore, \fazer finds most vulnerabilities among all fuzzers in 18 of the benchmarks, and in 11 benchmark programs finds unique vulnerabilities.  This provides evidence that \codename generalizes well across multiple programs and according to multiple metrics. 
We have released our fuzzer publicly and open-source~\cite{sivo-code}.

%% file: chapters/problem2.tex

\section{Problem}

Fuzzers look for inputs that trigger bugs in target programs. As the distribution of bugs in programs is unknown, fuzzers try to increase the chance of finding bugs by constructing inputs that 
lead to maximal program code execution. 
The objective of fuzzers is thus to construct inputs, called \emph{seeds}, that increase the amount of executed program code, called \emph{code coverage}. 
The coverage is measured based on the control-flow graph of the executed program, where nodes correspond to basic blocks (sets of program statements) and edges exist between sequential blocks. Some of the nodes are conditional (e.g. correspond to \textsf{if} and \textsf{switch} statements) and have multiple outgoing edges. Coverage increases when at some conditional node, called a \emph{branch}, the control flow takes a new edge which is not seen in previous tests---this is called \emph{inverting} or \emph{flipping} a branch. 

Grey-box fuzzers assess code coverage by instrumenting the programs and profiling coverage data during the execution of the program on the provided inputs. They maintain a pool of seeds that increase coverage. 
A grey-box fuzzer selects one seed from its pool, applies to it different operations called \emph{mutations} to produce a new seed, and then executes the program on the new seed. 
Those new seeds that lead to previously unseen coverage are added to the pool.
%
%
To specify a grey-box fuzzer one needs to define its seed selection, the types of mutations it uses, and the type of coverage it relies on. 
All these fuzzing components, we call \emph{stages}  or \emph{subroutines} of grey boxes.  We consider a few research questions related to different stages of fuzzing. 

\paragraph{RQ1: Impact of Complete Parameterization?} Fuzzers optimize for coverage.  There is no single fuzzing strategy that is expected to work well across all programs. So, the use of multiple strategies and optimization seems natural. 
Existing fuzzers do use dynamic strategy selection and optimize the parameter value selection.
For example, MOpt~\cite{lyu2019mopt}, AFLFast~\cite{bohme2017coverage}, and EcoFuzz~\cite{yue2020ecofuzz} use optimization techniques for
input seed selection and mutations. But, often such parameterization comes with 
internal constants, which have been hand-tuned on certain programs, and it is almost
never applied universally in prior fuzzers. The first question we ask is what would be the result of complete parameterization, i.e., if we encode all subroutines and their built-in constants as
optimization parameters.


The problem of increasing coverage is equivalent to the problem of inverting more branches. In the initial stage of fuzzing, when the number of not yet inverted branches is high, AFL mutation strategies (such as mutation of randomly chosen bytes) are successful and often help to invert branches in bulk. However, easily invertible branches soon become exhausted, and different strategies are required to keep the branch inversion going. One way is to resort to targeted inversion. In targeted inversion, the fuzzer chooses a branch and mutates input bytes that influence it. The following two questions are about refining target inversion in grey-box fuzzing.

\paragraph{RQ2: Efficient Taint Inference?}
Several fuzzers have shown that taint information, which identifies input bytes that influence a given variable, is useful to targeted branch inversion~\cite{rawat2017vuzzer,choi2019grey,you2019profuzzer,chen2018angora,aschermann2019redqueen,gan2020greyone}. If we want to flip a particular branch, the input bytes on which the branch condition variables depend should be mutated while keeping the other bytes unchanged. The main challenge, however, is to efficiently calculate the taint information.
Classical methods for dynamic taint-tracking incur significant instrumentation overheads whereas static methods have false negatives, i.e. they miss dependencies due to imprecision. The state-of-the-art fuzzers aim for light-weight techniques for dynamically inferring taint during fuzzing itself. Prior works have proposed methods which require number of tests linear in $n$, the size of the seed input~\cite{gan2020greyone}. This is extremely inefficient for programs with large inputs. This leads to our second question: Can we compute useful taint information but with exponentially fewer tests?

\paragraph{RQ3: Efficient Constraint-based Reasoning?}
Taint only captures whether a change in certain values of an input byte may lead to a change in the value of a variable. If we are willing to compute more expressive symbolic constraints, determining the specific input values which cause a program branch to flip is possible.
The challenge is that computing and solving expressive constraints, for instance first-order SAT/SMT symbolic formulae, is computationally expensive. In this work, we ask: Which symbolic constraints can be cheap to infer and solve during grey-box fuzzing?

\paragraph{RQ4: Precise coverage measurement?}
Grey-box fuzzers use coverage information as feedback to guide input generation. AFL, and almost all other fuzzers building on it, use control-flow edge counts as a common metric. Since there can be many
control-flow edges in the program, space-efficient data structures for storing runtime coverage data are important. Recent works have pointed out AFL's hash-based coverage map can result in collisions~\cite{gan2018collafl}, which has an unpredictable impact on the resulting optimization. 
How do we compute compressed edge counts with high precision using standard compilers for instrumentation?

\section{Overview of \codename}
%

Grey-box fuzzers instrument the target program to gather runtime
profiling data, which in turn guides their seed generation strategies. 
The objective of \fazer is to generate seeds that increase code coverage by using better and more of the profiling data. 
\fazer addresses the four research questions with four refinements. 


\paragraph{Parametrize-optimize approach (RQ1).}
\codename builds on the idea of complete parameterization of all fuzzing subroutines and strategies, i.e.  
none of the internal parameters are hard-coded. \codename selects strategies
and parameter values dynamically based on the observed coverage statistics, using a standard optimization algorithm. 
Such complete parameterization and optimization inherently makes \codename adaptable
to the target program and more general, since specialized strategies that work best for the program
are prioritized. To answer RQ1, we empirically show in our evaluation that this design principle individually helps \codename outperform other evaluated fuzzers across multiple target programs.

\paragraph{Fast Approximate Taint Inference (RQ2).}  We devise a fast and approximate taint inference engine \taint based on probabilistic group testing~\cite{du2000combinatorial}. 
Instead of testing individually for each input byte, \taint tests for carefully chosen groups of bytes and then combines the results of all tests to infer the taint for each individual byte.
This helps to reduce the test complexity of taint inference from $O(n)$ to $O(\log n)$ executions of the program, where $n$ is the number of input bytes. Thus the fuzzer can infer useful taint dependency even for very large inputs using \taint. 

\paragraph{Symbolic Interval Constraints (RQ3).}  
We propose inferring symbolic interval constraints that capture the relationship between inputs and variables used in branch conditions only. Instead of deductively analyzing the semantics of executed
instructions, we take an optimistic approach and infer these constraints from the observed values
of the inputs and branch conditional variables. The value-based inference is computationally cheap and tailored for a common case where values of the variables are direct copies of the inputs and when branches have comparison operations ($=,\neq, <,\le,>,\ge$). We show that such a constraint system can be solved efficiently as well without the use of SAT / SMT solvers. 

\paragraph{Compressed and Precise Edge Count Recording (RQ4).} 
We tackle both the collision problem and the compressed edge count problem in tracking coverage efficiently during grey-box fuzzing. 
For the former, we show a simple strategy based on using multiple basic block labels (rather than only one as in AFL) and reduce or entirely eliminate the collisions. For the later, to improve the prospect of storing important edge counts we propose temporary coverage flushing (i.e. resetting the coverage to zero). Although this may appear
to be a minor refinement in grey-box fuzzing, we find that it has a noticeable impact experimentally.

%% file: chapters/refinements.tex
\section{Design}
\label{sect:refinements}

We present the details of our four refinements in Sections~\ref{sec:learning}-\ref{sec:refcoverage} and then show the complete design of \fazer in Section~\ref{sect:design}.

\input{chapters/refine-learning}

\input{chapters/refine-dependency-inference}
\input{chapters/refine-system-interval}
\input{chapters/refine-coverage}

%% file: chapters/refine-learning.tex

\subsection{The Complete Parameterization Paradigm}
\label{sec:learning}

The \codename grey-box fuzzer aims to increase the code coverage in the fuzzed programs. 
Two points are central to this goal.
First, fuzzed programs come in different flavors, hence the fuzzer should be flexible and adaptive. 
We tackle the first point with \emph{parametrization}, i.e. by expanding the choice of available fuzzer subroutines. 
Second, a fuzzer has a few stages (i.e., selection of seeds, choice of mutations and their parameters, etc), and each one of them can be optimized. 
%
To address this point, we apply a complete \emph{optimization} of all available parameters. 


\vspace{10pt}
\paragraph{Parametrization.} 
The more fuzzing subroutines are available, the higher the chance that some of them may be optimal for fuzzing the targeted program.
Thus it is useful to expand the set of available fuzzing subroutines. To do so, we:
\begin{itemize}
\item {\em Add many fuzzing subroutines}.  For instance,  in addition to the AFL-style {\em vanilla} mutations that do not require any dependency information (e.g. mutate random bytes), we implement \emph{data-flow} strategies that utilize input dependency of program branches (e.g, mutation of dependent bytes). Besides adding new mutations, we also add more seed prioritization methods that determine how to sample a seed from the pool. 
\item {\em Introduce variations in each subroutines}. Often this can be done by varying internal hard-coded parameters in subroutines. For instance, in the mutation of random bytes, instead of changing a single byte, \codename can change 1, 2, 4, 8, 16, 32, or 64 bytes at once. The exact number of bytes is considered an input parameter; it can take one of the above 7 values (and the choice of value potentially can be optimized). Not all variations in subroutines are effected with changing integer parameters. For instance, the seed selection criterion is based on speed, number of repetitions, length of seed, and so on. These variations are enumerated and serve as an input parameter to the seed criterion. All such parameters to subroutines are optimized per program.

\end{itemize}
As a result, across the whole fuzzer, there are 17 different fuzzing subroutines with 68 variations.  In comparison, the baseline AFL has around 15 different subroutines with around 45 variations\footnote{Despite having comparable numbers, \fazer and AFL use mostly different mutations and thus subroutines. }.

\vspace{10pt}
\paragraph{Optimization.} 
The parametrization increases the chance that potentially optimal subroutines are chosen for each program. The next step is to select which subroutines are turned on for a given program. 
It is critical to understand that we are not dealing with a single optimization problem. 
Fuzzing is a continuous process, composed of iterations that select a seed and a mutation, apply the mutation to the seed, and check on coverage increase. Thus, in each iteration we need to optimize the selection of fuzzing subroutines several times---for example, the used seed criterion and class, the mutation strategy, (potentially a number of) mutations sub-strategies, the inputs to the mutation strategy, and so on.  
For this purpose, we use multi armed bandits (MAB), a simple reinforcement learning algorithm. 
Given a set of choices, each choice providing a certain reward when selected, MAB helps to select the choices such that their accumulative rewards are maximized. The rewards are unknown and stochastic, and the selection process is continuous. Note, after MAB selects a choice, it needs to receive as a feedback the obtained reward to update its choice selection strategy.    

Reducing the selection of fuzzing subroutines to MAB problem is straightforward. First, note that we consider each selection as an independent MAB problem, for instance, the optimal number of random bytes to mutate is one MAB problem. 
Our objective is to maximize the coverage, hence it is natural to use the additional coverage acquired from executing the choice as the MAB {\em reward}. However, this metric alone may not be accurate because some choices incur higher computational costs. Therefore, we use the \emph{additional coverage per time unit} as the reward. 
 In the conventional MAB, the distributions of rewards are stationary with some unknown mean. In our case, as the fuzzer progresses, it requires more computational effort to reach the remaining  unexplored code and increase coverage. In other words, the rewards for the selection choices monotonically decrease over time. 
Therefore, we model our problem as MAB with \emph{non-stationary} rewards and use discounting to solve it~\cite{kocsis2006discounted}. 
For more details on application of MAB in \fazer, we refer the reader to  Algorithm~\ref{alg:iteration} and Section~\ref{sect:design}.

Besides optimizing for selections, during fuzzing we use genetic algorithms (GA) to optimize any black-box objective functions that arise during  inversion of branches. More precisely, we reduce parts of the inversion problem to a black-box optimization and apply GA to speed-up the inversion.  
In vanilla type fuzzing, we use GA to search for optimal positions of a fixed number of bytes to mutate. For this purpose, as an objective function we use the number of branches that are affected (i.e. change some of their variables) when selected bytes are mutated.  
As the number of affected branches increases, so does the chance of inverting one of them by mutating the found optimal position bytes. 
In data-flow fuzzing, with GA  we invert non-discriminatory targeted branches. In this case, as an objective function we use the distance of the resulting branch value to the value that corresponds to branch inversion. For instance, for the branch "\textsf{ if(x == 5) }", the objective function is $|x - 5|$. When the distances reaches zero, i.e. the minimum of the objective function is reached, the branch is inverted.

%% file: chapters/refine-dependency-inference.tex

\subsection{Fast Approximate Taint Inference}
\label{sec:inference}

\input{chapters/code-sample}
To infer dependency of branches on input bytes, earlier fuzzers relied on the truth value of branch conditions: if changing the value of a particular byte changes the truth value of a branch, then it is inferred that the branch depends on this byte.  
For instance, in Fig.~\ref{fig:branches}, to correctly infer the dependency of the branch at line~\ref{line:B1}, the engine first needs to select for mutation the input byte \texttt{x[100]} and then to change its value from any other than 40 to 40.
\go~\cite{gan2020greyone} proposed so-called \emph{fuzzing-driven taint inference} FTI by switching the focus from the truth value of a branch to the value of the variables used in the branch.
For instance, FTI determines the dependency of branch at line~\ref{line:B1} on \texttt{x[100]} as soon as this input bytes is mutated, because this will lead to a change of the value of the variable $A$ that is used in the branch.    
FTI is sound (no over-taint) and incomplete (some under-taint). 
Exact reasoning with provable soundness or completeness is not a direct concern in fuzzers, since they only use it to generate tests which are concretely run to exhibit bugs.

The prime issue with FTI, which improves significantly over many other prior data-flow based engines, is efficiency. The taint is inferred by mutating bytes one-by-one in FTI. Thus, to infer the full dependency on all input bytes, the engine will require as many executions as the number of bytes. A seed may have tens of KBs, and there may be thousands of seeds,  therefore the full inference may quickly become a major bottleneck in the fuzzer. On the other hand, precise or improved branch dependency may not significantly boost fuzzer bug-finding performance, thus long inference time may be unjustified. Hence, it is critical to reduce the inference time.

\paragraph{The \taint engine. }
We use \emph{probabilistic group testing}~\cite{du2000combinatorial} to reduce the required number of test executions for potential full inference from $O(n)$ to $O(\log n)$, where $n$ is the number of input bytes. 
Instead of  mutating each byte individually followed by program execution (and subsequent FTI check for each branch condition if any of its variables has changed), we simultaneously mutate multiple bytes, and then execute the program with the FTI check. We choose the mutation positions non-adaptively, according only to the value of $n$. This assures that dependency for many branches can be processed simultaneously. 

Consider the code fragment at Figure~\ref{fig:branches} (here $n=1024$). 
We begin the inference by constructing 1024-bit binary vectors $V_i$, where each bit corresponds to one of the input bytes. A bit at position $j$ is set iff the input byte $j$ is mutated (i.e. assigned a value other than the value that has in the seed). Once $V_i$ is built, we execute the program on the new input (that corresponds to $V_i$) and for each branch check if any of its variables changed value (in comparison to the values produced during the execution of the original seed). If so, we can conclude that the branch depends on some of the mutated bytes determined by $V_i$.  
Note, in all prior works, the vectors $V_i$ had a single set bit (only one mutated byte). As such, the inference is immediate, but slow.  
On the other hand, we use vectors with $\frac{1024}{2}=512$ set bits and select  $2\cdot \log_2 1024=20$ such vectors.  Vectors $V_{2\cdot j }, V_{2\cdot j + 1}$ have repeatedly $2^j$ set bits, followed by $2^j$ unset bits, but with different starts. For instances, the partial values of the first 5 vectors $V_i$ are given below on the right.
$$
\begin{matrix}
V_0 = 1010101010101010101010... \\ 
V_1 = 0101010101010101010101... \\
V_2 = 1100110011001100110011...\\
V_3 = 0011001100110011001100...\\
V_4 = 1111000011110000111100...\\
...\\
\end{matrix}
$$
We execute the resulting 20 inputs and for each branch  build 20-bit binary vector $Y$. The bit $i$ in $Y$ is set if any of the branch values changed after executing the input that corresponds to $V_i$.  For instance, for the branch at line~\ref{line:B1} of Figure~\ref{fig:branches}, $Y=10100110100101101010$.
Finally, we decode $Y$ to infer the dependency. To do so, we initialize 1024-bit vector $D$ that will hold the dependency of the branch on input bytes---bit $i$ is set if the branch depends on the input byte $i$. We set all bits of $D$, i.e. we start by guessing full dependency on all inputs. Then we remove the wrong guesses according to $Y$.  
For each unset bit $j$ in $Y$ (i.e. the branch value did not change when we mutated bytes $V_j$), we unset all bits in $D$ that are set in $V_j$ (i.e. the branch does not depend on any of the mutated bytes $V_j$). 

After processing all unset bits of $Y$, the vector $D$ will have set bits that correspond to potential dependent input bytes. 
Theoretically, there may be under and over-taint, according to the following information-theoretic argument: $Y$ has 20 bits of entropy and thus it can encode at most $2^{20}$ dependencies, whereas a branch may depend on any of the 1024 input bytes and thus it can have $2^{1024}$ different dependencies. In practice, however, it is reasonable to assume that most of the branches depend only on a few input bytes\footnote{C-type branches that contain multiple variables connected with AND/OR statements, during compilation are split into subsequent independent branches. Our inference is applied at assembly level, thus most of the branches depend only on a few variables.}, and in such a case the inference is more accurate. For branches that depend on a single byte, the correctness of the inference follows immediately from group testing theory\footnote{The matrix with rows $V_0,V_1,\ldots$ is $1$-disjunct and thus it can detect 1 dependency. 
}. 
For instance, the branch at line~\ref{line:B1} of Fig~\ref{fig:branches} will have correctly inferred dependency only on byte \texttt{x[100]}.
For branches that depend on a few bytes, we can reduce (or entirely prevent) over-taint by repeating the original procedure while permuting the vectors $V_i$. In such a case, each repeated inference will suggest different candidates, except the truly dependent bytes that will be suggested by all procedures. These input bytes then can  be detected by taking intersection of all the suggested candidates.  
For instance, for the branch at line~\ref{line:B2} (that actually depends on 8 bytes), a single execution of the procedure will return 16 byte candidates. By repeating once the procedure with randomly permuted positions of $V_i$, with high probability only the 8 actual candidates will remain.

The above inference procedure makes the implicit assumption that \emph{same branches are observed across different executions}. 
Otherwise, if a branch is not observed during some of the executions, then the corresponding bit in $Y$ will be undefined, thus no dependency information about the branch will be inferred from that execution. For some branches the assumption always holds (e.g. for branches at lines~\ref{line:B1},\ref{line:B2} in Figure~\ref{fig:branches}). 
For other branches, the assumption holds only with some probability that depends on their branch conditions. For instance, the branch at line~\ref{line:B4} may not be seen if the branch at line~\ref{line:B3} is inverted, thus any of the 20 bits of $Y$ may be undefined with a probability of $\frac{200}{2^{32}}$. 
In general, for any branch that lies below some preceding branches, the probability that bits in $Y$ will be defined is equivalent to the probability that none of the above branches will inverted by the mutations\footnote{This holds even in the case of FTI. However, the probabilities there are higher because there is a single mutated byte.}. As a rule of thumb,  the deeper the branch and the easier to invert the preceding branches are, the harder will be to infer the correct dependency. 
%
%
To infer deeper branches, we introduce a modification based on forced execution. 
%
We instrument the code so the executions at each branch will take a predefined control-flow edge, rather than decide on the edge according to the value of the branch condition.
This guarantees that the target branches seen during the execution of the original seed file (used as a baseline for mutation), will be seen at executions of all subsequent inputs produced by mutating the original seed. 
%
We perform forced execution dynamically, with the same statically instrumented program, working in two modes. In the first mode, the program is executed normally, and a trace of all  branches and their condition values is stored. In the second mode, during execution as the branches emerge, their condition values are changed to the stored values, thus the execution takes the same trace as before. No other variables aside from the condition values are changed. Note that our procedure aims to infer taint dependencies fast and optimistically; we refer readers to Section~\ref{sec:limitations} for a discussion on these aspects.

%% file: chapters/code-sample.tex
\begin{figure}[!h]
\centering
\begin{lstlisting}[style=CStyle]
//input is uint8_t x[1024]
A = x[100] + 10;				
B = (uint32_t *)x[200];  	
C = (uint32_t *)x[236]; 	
...
if ( A == 50)				@\label{line:B1}@
	...
if( B + C < 200 )			@\label{line:B2}@
	...
if( C  > 200 ) {					@\label{line:B3}@
	...
	if( A + C < 400 ) 				@\label{line:B4}@
		...
}

\end{lstlisting}
\captionof{figure}{Branches with dependent input bytes.}
\label{fig:branches} 
\end{figure}    

%% file: chapters/refine-system-interval.tex
\subsection{Solving System of Intervals}
\label{sec:systemintervals}

It was noted in RedQueen~\cite{aschermann2019redqueen}, that when branches depend trivially on input bytes (so-called direct copies of bytes) and the branch condition is in the form of equality (either $=$ or $\neq$), then such branches can be solved trivially. For instance, the branch at line~\ref{int:line:B1} of 
Figure~\ref{fig:intervals}, depends trivially on the byte \texttt{x[0]} and its condition can be satisfied by assigning $x[0]=5$ (or inverted by assigning $x[0]\neq5$). 
\input{chapters/code-interval}
Thus it is easy to satisfy or invert such branches, as long as the dependency is correctly inferred and the branch condition is equality. 
Similar reasoning, however, can be applied when the condition is in the form of inequality over integers. Consider the branch at line~\ref{int:line:B2} of Figure~\ref{fig:intervals}, that depends trivially on the input byte $x[1]$. From the type of inequality (which can be obtained from the instruction code of the branch), and the correct dependency on the input byte $x[1]$ and the constant $100$, we can deduce the branch form  $x[1]<100$, and then either satisfy it resulting in  $x[1] \in [0,99]$, or invert it, resulting in $x[1] \in [100,255]$. In short, we can represent the solution in the form of integer intervals for that particular input byte. 

Often to satisfy/invert a branch we need to take into account not one, but several conditions that correspond to some of the branches that have common variables with the target branch. For instance, to satisfy the branch at line~\ref{int:line:B4}, we have two inequalities and thus two intervals: 
$x[2]\in[0,200]$ corresponding to target branch at line~\ref{int:line:B4} and $x[2]\in[11,255]$ corresponding to branch at line~\ref{int:line:B3}. Both share the same input variable $x[2]$ with the target branch. A solution ($x[2]\in [11,200]$) exists because the intersection of the intervals is not empty.  

In general, \codename builds a system of such constraints starting from the target branch, by adding gradually preceding branches that have common input variables with the target branch. Each branch (in)equality is solved independently immediately, resulting in one or two intervals (two intervals only when solving $x \neq value$, i.e. $x\in [0,value-1]\cup[value+1, maxvalue]$), and then intersection is found with the previous set of intervals corresponding to those particular input bytes. 
Keeping intervals sorted assures that the intersection will be found fast.
Also, each individual intersection can increase the number of intervals at most by 4. Thus the whole procedure is linear 
in the number of branches along the executed path. As a result, we can efficiently solve these type of constraints and, thus, satisfy or invert branches that depend trivially on input bytes. 

Even when some of the preceding branches do not depend trivially on input bytes, solving the constraints for the remaining branches gives an advantage in inverting the target branch. In such a case, we repeatedly sample solutions from the solved constraints and expect that the non-inverted branch constraints will be satisfied by chance. 
As sampling from the system requires constant time (after solving it), the complexity of branch inversion
is reduced only to that of satisfying non-trivially dependent branches. 
For instance, to reach line~\ref{int:line:B6}, we first solve the lines~\ref{int:line:B3},~\ref{int:line:B4} to obtain  $x[2]\in[11,200]$, and then keep sampling $x[2]$ from this interval and hope to satisfy the branch at line~\ref{int:line:B5} by chance.


%% file: chapters/code-interval.tex

\begin{figure}[!h]
\centering

\begin{lstlisting}[style=CStyle]
if ( x[0] == 5 )			@\label{int:line:B1}@
	...
if( x[1] < 100 )	@\label{int:line:B2}@
	...
if( x[2]  > 10 ) {					@\label{int:line:B3}@
	...
	if( x[2] <=  200 ) 			@\label{int:line:B4}@
		...
		if( foo(x[2]) ==  0 ) 			@\label{int:line:B5}@
			...					@\label{int:line:B6}@
}

\end{lstlisting}
\label{fig:intervals} 
\captionof{figure}{Branches and systems of intervals.}
\end{figure}    

%% file: chapters/refine-coverage.tex

\subsection{More Accurate Coverage}
\label{sec:refcoverage}

AFL uses a simple and an elegant method to record the edges and their counts by using an array \textsf{showmap}. First, it instruments all basic blocks $B_i$ of a program by assigning them a unique random label $L_i$. Then, during the execution of the program on a seed, as any two adjacent basic blocks $B_j,B_k$ are processed, it computes a hash of the edge $(B_j,B_k)$ as $E=(L_j\ll1) \oplus L_k$ and performs \textsf{showmap[E]++}. New coverage is observed if the value $\lfloor \log_2 \textsf{showmap[E]} \rfloor $ of a non-zero entry \textsf{showmap[E]} has not been seen before. If so, AFL updates its coverage information to include the new value, which we will refer to as the logarithmic count. 

\paragraph{Prevent colliding edge hashes.}
CollAFL~\cite{gan2018collafl} points out that when the number of edges is high, their hashes will start to collide due to birthday paradox, and  \textsf{showmap} will not be able to signal all distinct edges. Therefore, a fuzzer will fail to detect some of the coverage. 
%
%
We propose a simple solution to the collision problem. Instead of assigning only one label $L_i$ to each basic block $B_i$, we assign several labels $L_i^1,\ldots,L_i^m$, but use only one of them during an execution. The index of the used label is switched occasionally for all blocks simultaneously. The  switch assures that with a high chance, each edge will not collide with any other edge at least for some of the indices. The number of labels required to guarantee that all edges will be unique with a high chance at some switch depends on the number of edges. Due to space restrictions we omit the combinatorial analysis. 
We provide a combinatorial analysis in the Appendix~\ref{sect:appendix:multibloc} on this number. 
In our actual implementation the size of the \textsf{showmap} is $2^{16}$ and we use $m=4$ labels per basic block -- on average this allows around 8,000 edges to be mapped uniquely (and even 20,000 with less than 100 collisions), which is sufficiently high quantity for most of the programs considered in our experiments. By default, the index is switched once every 20 minutes.

\paragraph{Improve compressed edge counts.} The logarithmic count helps to reduce storing all possible edge counts, but it may also implicitly hinder achieving better coverage.  This is because
certain important count statistics that have the same logarithmic count as previously observed during fuzzing might be discarded. 

\input{chapters/code-compressed-path}
For instance, if the \textsf{for} loop in Figure~\ref{fig:aflcompress} gets executed 13 times, then AFL will detect this as a new logarithmic count of  $\lfloor \log_2 \textsf{13} \rfloor = 3$, it will update the coverage, save the seed in the pool, and later when processing this seed, the code block $F1()$ will be executed as soon as the condition $C1$ holds. On the other hand, afterwards if the \textsf{for} loop gets executed 14 times, then the same logarithmic count $\lfloor \log_2 \textsf{14} \rfloor = 3$ is achieved, thus the new seed will not be stored, therefore the chance of executing the code block $F2()$ is much lower. In other words, to reach $F2()$, simultaneously the \textsf{for} loop needs to be executed 14 times and $C2$ condition needs to hold. Hence, $F1()$ and $F2()$ cannot be reached with the same ease despite having similar conditional dependency, only because of AFL's logarithmic count mechanism.

To avoid this issue, we propose flushing the coverage information periodically. 
More precisely, periodically we store the current coverage information, then reset it to zero, and during some time generate new coverage from scratch. After exhausting the time budget on new coverage, we keep only the seeds that increase the stored coverage, and continue the fuzzing with the accumulated coverage.

%% file: chapters/code-compressed-path.tex
\begin{figure}[!h]
\centering
\begin{lstlisting}[style=CStyle]
count = 0;
for(i=0; i< x; i++)
	count++;
	
if( 13 == count && C1 )
	F1();		@\label{line:F1}@
else if( 14 == count && C2 )
	F2();		@\label{line:F2}@
	
\end{lstlisting}
\caption{The effects of AFL's edge count compression.}
\label{fig:aflcompress} 
\end{figure}    

%% file: chapters/design.tex

\subsection{Design  of the Whole Fuzzer \fazer}
\label{sect:design}

\input{chapters/design-table}

\fazer implements all the refinements mentioned so far. It uses the standard grey-box approach of processing seeds iteratively. In each iteration, it selects a seed, mutates it to obtain new seeds, and stores those that increase coverage. 


\begin{algorithm}[h]
use\_class  		$\gets$ MAB\_select( Seed\_class )		\tcp*{choose seed class with MAB} 
use\_crit 		$\gets$ MAB\_select( Seed\_criterion ) 		\tcp*{choose seed criterion} 
seed			$\gets$ Sample( use\_class , use\_crit, Seeds )	\tcp*{sample seed from the pool}
use\_strategy 		$\gets$ MAB\_select( Fuzzer\_strategy ) \tcp*{choose Data-flow or Vanilla} 
\If{ use\_strategy == Data-flow}{						
	Taint\_inference(seed)								\tcp*{if Data-flow then infer dependency}
}
tot\_cov\_incr $\gets$  0\;
\While{time budget left }{
	use\_mut $\gets$ MAB\_select( strategy )		\tcp*{choose one mutation}
	use\_mut\_params $\gets$ MAB\_select( use\_mut )	\tcp*{choose its params}	
	new\_seed $\gets$ Mutate( seed, use\_mut, use\_mut\_params ) \tcp*{apply mutation}
	new\_coverage $\gets$ ProduceCoverage(new\_seed)\;
	cov\_increase $\gets$ $\|$ new\_coverage $\setminus$ Coverage $\|$ \tcp*{new coverage?}
	\If{ cov\_increase $>$ 0  }{		
		Seeds $\gets$ Seeds $\bigcup$ new\_seed  	\tcp*{add new seed to the pool}
		Coverage $\gets$ Coverage $\bigcup$ new\_coverage  \tcp*{update coverage}
	}
	\tcp*{feedback cov/sec to MAB to update the effectiveness of the chosen mutation and its params }
	MAB\_update( [use\_mut , use\_mut\_params], cov\_increase, while\_time )\;
	tot\_cov\_incr += cov\_increase \;	  
}
\tcp*{feedback total cov/sec to MAB to update the effectiveness of the chosen seed class/criterion and fuzzing strategy }
MAB\_update( [use\_class,use\_crit,use\_strategy] , tot\_cov\_incr  , iter\_time)\;
\caption{OneIteration\fazer( Seeds, Coverage )}
\label{alg:iteration}
\end{algorithm}
\begin{algorithm}[h]
Seeds  		$\gets$ Initial\_seeds\;
Coverage $\gets$ ProduceCoverage(Seeds)\;
\While{ true }{

	OneIteration\fazer( Seeds, Coverage );
	
	\If{ time\_to\_switch\_index }{
		SwitchIndexInCoverage()\;
		Coverage $\gets$ ProduceCoverage( Seeds )\;
	}

	\If{ time\_to\_start\_flush }{
		Old\_coverage, Old\_seeds $\gets$ Coverage, Seeds\;
		Seeds $\gets $ Initial\_seeds\;
		Coverage $\gets$ ProduceCoverage( Seeds )\;
	}
	
	\If{ time\_to\_stop\_flush }{
		New\_coverage $\gets$ Coverage $\setminus$ Old\_coverage\;
		Coverage $\gets$ Coverage $\bigcup$ Old\_covarege\;
		Seeds $\gets$ Old\_seeds  $\bigcup$  GetSeedsThatProduceCov(Seeds, New\_coverage) \;
	}

}
\caption{\fazer}
\label{alg:fuzzer}
\end{algorithm}


In \fazer (refer to the pseudo-code in Algorithm~\ref{alg:iteration}), the seed selection is optimized: first with MAB the currently best class and best criterion are selected, and then a seed is sampled from the pool according to the chosen class and criterion. 
Afterwards, the fuzzer with the help of MAB decides on the currently optimal fuzzing strategy, either vanilla (apply mutations that do not require dependency information) or data-flow (require dependency). If latter, \fazer first infers the dependency (as a combination of FTI and \taint).   
Then, according to the chosen fuzzing strategy the fuzzer again uses MAB to select one optimal mutation strategy. The vanilla fuzzing strategy allows a choice of 3 different mutations: 1) mutation of random bytes, 2) copy/remove of byte sequence of current seed, and 3)  concatenation of different seeds. On the other hand, data-flow fuzzing strategy consists of 5 mutations: 1) mutation of dependent bytes, 2) branch inversion with system solver, 3)branch inversion by minimizing objective function, 4) branch inversion by mutation of their dependent bytes, and 5) reusing previously found bytes from other seeds to current seed. 
Most mutations have sub-versions  or parameters which are also chosen with MAB. For instance, mutation of random bytes supports two versions: it can use heuristics to determine the positions of the bytes (choice 1), or use random byte positions (choice 2). If choice 1, then it needs to select the number of mutated bytes ($1,2,4,8,16,32,$ or $64$). Both of these selections are determined with MAB. 
Each mutation is applied to the chosen seed to obtain a new seed, and then  the seed is executed.  
The coverage update information is fed back to the MAB, thus assuring that MAB can further optimize the selections. 
In Table~\ref{tbl:mab} we give a  list of selections available at different steps of the fuzzer.  

\fazer runs the iterations and occasionally executes the code coverage refinements -- refer to Algorithm~\ref{alg:fuzzer}. 
To understand the additional parts introduced by the refinements of \fazer, in Appendix~\ref{sect:appendix:code} we provide a pseudo-code of a generic grey-box fuzzer.

%% file: chapters/design-table.tex

\begin{table*}[!ht]
  \centering
    \caption{Fuzzer procedures and their variations}

  \begin{tabular}{lll}
 Procedure/Parameters & Variation(s) & Description \\
 \hline 
 \hline
 \textsf{Seed\_class}  	& SC-fast-edges 			& consider only most efficient seeds for each edge  \\
 					& SC-fast-multiple-edges 	& include as well the fastest for each multiplicative edge  \\
 					& SC-all 				& consider all seeds  \\
 \hline
 \textsf{Seed\_criterion}  & Count  & choose the least sampled seed  \\
 					& Speed  & sample according to number of executions per second \\
 					& Length & sample according to number of bytes  \\
 					& Crash   & consider only seeds that lead to crashes  \\
 					& Cov      & consider only seeds that increase edge count  \\
 					& Random& sample randomly   \\
 \hline
 \textsf{Fuzzer\_strategy} & Vanilla    & does not require taint inference \\
					& Data-flow & requires taint inference \\
 \hline
 \textsf{Vanilla} 		& Mutate-rand-bytes    	& mutate random bytes \\ 
					& Copy-remove		& copy and remove byte sequences of current seed \\
					& Combiner			& combine multiple seeds at different positions\\
 \hline
 \textsf{Data-flow} 		& Mutate-bytes		& mutate dependent bytes \\
 			 		& Invert-branches    	& invert target branches by mutating their dependent bytes \\ 
 			 		& Invert-branches-GA 	& invert target branches with GA by minimizing objective function \\ 
 			 		& System-solver		& invert branches with system solver   \\ 
 			 		& Mingler 			& reuse previously found bytes from other seeds to this seed  \\ 					
 \hline
 \textsf{Mutate-rand-bytes/Type} 		& MRB-GA		& use GA to determine positions of mutated bytes  \\
 			 						& MRB-simple    	& mutate randomly chosen bytes \\ 
 \hline
 \textsf{MRB-GA} 						& 1,2,4,8,16,32,64		& number of mutated bytes   \\
 \hline
 \textsf{MRB-simple } 					& True,False			& bias selection of bytes according to their previous use   \\
 \hline
 \textsf{Copy-remove/Number} 			& 1,2,4,8,16,32,64		& the number of copy/remove operations\\
 \hline
 \textsf{Copy-remove/Mode} 			& CR-rand			& add random bytes/remove \\
 			 						& CR-real				& copy real bytes/remove \\
 			 						& CR-learn			& learn and use byte divisors seen previously 	\\
 			 						& CR-prev			& copy/remove at positions previously successfull	\\
 \hline
 \textsf{Combiner/Number} 				& 2,3,4,5,6,7,8		& the number of different seeds to combine \\
\hline
 \textsf{Combiner/Select} 					& Speed/Inverse-speed & Prefer seeds that are faster/slower to execute \\
 			 						& Length/Inverse-length& Prefer seeds that are shorter/longer  \\
 			 						& Select-random		& Sample randomly  	\\
\hline									
 \textsf{Combiner/Mode} 					& CM-random			& combine at random positions  \\
 									& CM-learn				& learn and use byte divisors to select position \\
\hline									
 \textsf{Mutate-bytes/Number} 			& 1,2,4,8,16			& number of dependent bytes to mutate at once  \\
\hline									
 \textsf{Mutate-bytes/Type} 				& True,False			& bias selection of bytes according to their previous use   \\
 \hline
 \textsf{System-solver/Type} 					& ST-all			& add at once all branches dependent on the target inversion branch \\
 										& ST-one			& add gradually one by one the unsolved branches\\
\hline									
 \textsf{Mingler/Number} 				& 1,2,3,4,5,6				& number of previous byte solutions to apply at once \\
 			 						  
\end{tabular}
  \label{tbl:mab}
\end{table*}

%% file: chapters/implementation.tex

\subsection{Implementation}
\label{sec:implementation}

We implement \fazer in C++ with around 20,000 lines of code.  All of the code is written from scratch, with the exception of around 500 lines related to so-called \emph{fork server}\footnote{The fork server helps to speed-up execution of programs. It runs once the initialization of system resources (required by \emph{execve}), stores the state, and all subsequent executions of the program are ran starting from the stored state.}, which is based on AFL's code. 
\fazer uses static instrumentation to obtain the data about the coverage and branches of the programs. More precisely, 
three of the four refinements require additional instrumentation of programs to implement their functionality: the accurate code coverage refinement requires lighter instrumentation, whereas \taint and system solver refinements require heavier instrumentation. 
For this purpose, we compile a program source code with two different LLVM passes: one that utilizes relatively lightweight instrumentation used only for code coverage, and one with heavier instrumentation that provides information about the branches as well. The compilation is done by a modified version of Clang (in a fashion similar to AFL's \emph{afl-clang++}). 
The overheads introduced by the instrumentation are given in Appendix~\ref{sect:appendix:instrumentation}.

%% file: chapters/refine-limitations.tex

\subsection{ Limitations of \fazer}
\label{sec:limitations}

 
The taint engine \taint relies on forced execution, which by definition is not sound, thus the inference is approximate. It means, the engine may introduce false positives/negatives, i.e. it may suggest dependencies of branches on incorrect input bytes.
This, however, is not a real concern in fuzzing because later it leads solely to mutating incorrect input bytes, hence potentially it has only impact on efficiency\footnote{The impact can be reduced with various methods, e.g., the MAB-based optimization presented in this paper. }, and does not affect the correctness of the fuzzer in any other way. 
The accuracy of the engine varies between programs. In certain cases (of particular traces), 
the forced execution crashes the program, and thus the inference has lower accuracy (because the corresponding $Y$ bit is undefined). In our actual implementation of \taint, we prevent some of the crashes by detecting with binary search sequences of input bytes that lead to crashes, and later eliminate them from consideration. 

The refinement based on system of intervals is neither sound nor complete. Problems may appear due to incorrect inference of the intervals as well as due to the fact that the system describes only a partial dependency of the target branch on input bytes, i.e. includes only branches that can be presented in the form of integer intervals. Therefore, one may not assume that all of the branches can be properly inverted using this refinement.     

The remaining two refinements do not have apparent limitations, aside from affecting the efficiency in some cases.  

%% file: chapters/eval.tex

\section{Evaluation}
\label{sec:evaluation}

We show that \fazer performs well on multiple benchmarks according to the standard fuzzing metrics such as code coverage (Section~\ref{sec:eval:coverage}) and found vulnerabilities (Section~\ref{sec:eval:vulnerabilities}). We evaluate the performance of each refinement in Section~\ref{sec:eval:stats}.

\input{chapters/eval-1-setup}

\input{chapters/eval-2-coverage}
\input{chapters/eval-3-vulnerabilities}

\input{chapters/eval-4-stats}

\input{chapters/eval-5-explain}

%% file: chapters/eval-1-setup.tex

\subsection{Experimental setup}
\label{sec:eval:setup}

\paragraph{Experiment environment.}
For all experiments we use the same box with 
Ubuntu Desktop 16.04, two Intel Xeon E5-2680v2 CPUs @2.80GHz with 40 cores, 64GB
DDR3 RAM @1866MHz and SSD storage. 
All fuzzers are tested on the same programs, provided with 
only one initial seed, randomly selected from samples available on the internet.
To keep experiments computationally reasonable, while still providing a fair
comparison of all considered fuzzers, we performed a two-round tournament-like assessment. 
In the first round, all fuzzers had been appraised 
over the course of 12 hours. This interval is chosen
based on Google's FuzzBench periodical reports, 
which shows that $12$ hours is sufficient to decide the ranking of the fuzzers usually~\cite{ossfuzz}.
%
The top 3 fuzzers from the first round that perform the best on average over all evaluated programs progress to the second round, in which they are run for 48 hours. 


\paragraph{Baseline fuzzers.}
We evaluate \fazer in relation to $11$ notable  grey-box fuzzers. In addition
to AFL~\cite{afl}, we take the extended and improved AFL family: AFLFast~\cite{bohme2017coverage}, 
FairFuzz~\cite{lemieux2018fairfuzz}, LAF-Intel~\cite{lafintel}, MOpt~\cite{lyu2019mopt} and EcoFuzz~\cite{yue2020ecofuzz}. 
Moreover, we include Angora~\cite{chen2018angora} for its unique mutation
techniques, Ankou~\cite{manes2020ankou} for its fitness function, 
and a few fuzzers that perform well on Google's OSS-Fuzz~\cite{ossfuzz} platform such as 
Honggfuzz~\cite{honggfuzz}, AFL++ and AFL++\_mmopt~\cite{fioraldi2020afl++} (version 2.67c).
%
To prevent unfair comparison, we omit from our experiments two categories of fuzzers. \
First, we exclude popular grey-box fuzzers that do not have an officially available implementation, such as CollAFL~\cite{gan2018collafl}  and  GreyOne~\cite{gan2020greyone}. We did not implement these fuzzers from scratch due to the complexity of such a task (e.g. the authors of GreyOne report 20K LoC implementation). Second, we exclude hybrid fuzzers because their approach is basically orthogonal to traditional grey-box fuzzers and thus they can be combined. For instance, the well-known hybrid fuzzer QSYM~\cite{yun2018qsym} inverts branches with symbolic execution and is built on top of AFL. With minor modification, QSYM could be built on top of \fazer instead of AFL, and this hybrid may lead to an even better performance.

\paragraph{Programs.}
Our choice of programs was influenced by multiple factors, such as 
implementation robustness, diversity of functionality, and previous
analysis in other works. 
Our main goal of the evaluation is comparison of fuzzers according to a few criteria (including discovery of bugs), thus we use versions of programs that have already been tested in prior fuzzer evaluations on similar criteria. Due to limited resources, we did not run the fuzzers on the latest versions to look for actual CVEs.
%
Our final selection consists of 27 programs including: binutils (e.g.:
\textsf{readelf}, \textsf{nm}), parsers and parser generators (e.g.:
\textsf{bson\_to\_json@libbson}, \textsf{bison}), a wide variety of analysis
tools (e.g.: \textsf{tcpdump}, \textsf{exiv2}, \textsf{cflow},
\textsf{sndfile-info@libsndfile}),
image processors (e.g.: \textsf{img2txt}), assemblers and
compilers (e.g.: \textsf{nasm}, \textsf{tic@libncurses}), compression tools
(e.g.: \textsf{djpeg}, \textsf{bsdtar}), the LAVA-M dataset \cite{lava-m}, 
etc.
A complete list of the programs, their version under test, and their input parameters  is available in Appendix~\ref{sect:appendix:programs}.

\paragraph{Efficiency metrics.}
We use two metrics to compare the efficiency of fuzzers: edge coverage and the number of found vulnerabilities. 
To determine the coverage, we use the logarithmic edge count because this number is the objective in the fuzzing routines of  the AFL family of fuzzers ~\footnote{In AFL, this is the number of set bits in the accumulative array \textsf{showmap}.}. For completeness, in Appendix~\ref{sect:appendix:coverage} we provide as well a simple edge count.  
The coverage data may be inaccurate due to colliding edges when measured as in AFL. To rectify this, we instrument the target programs to store the full execution traces and compute precise coverage data. In Appendix~\ref{sect:appendix:coverage} we also show the imprecise coverage, measured as in AFL. 

We measure the total number of distinct vulnerabilities found by each fuzzer. For this purpose, first we confirm the reported vulnerabilities, i.e. we take all seeds generated by a fuzzer and keep those that trigger a crash by any of the sanitizers ASAN~\cite{asan}, UBSAN~\cite{ubsan} and Valgrind~\cite{valgrind}.\footnote{ASAN and UBSAN detect more types of vulnerabilities than Valgrind, however, they also require additional instrumentation during the compilation of the programs. We use ASAN and UBSAN to confirm the vulnerabilities in such programs. For the programs that fail to compile with the additional instrumentation, we use Valgrind as an alternative. This criterion was used for all fuzzers.} 
Then, for each kept seed, we record the source line in the program that triggers a crash (according to the appropriate sanitizer). Finally, we count the number of found distinct source lines for the crash point. 

\vspace{5pt}
\noindent
\textit{Remark.} Some fuzzers~\cite{gan2020greyone,gan2018collafl} report so-called path coverage measured as the number of produced seeds. We find this metric rather inaccurate. For instance, a seed that inverts $n$ branches often is equivalent to $n$ seeds that invert a single branch. However, the above path coverage will consider the former as a single path, while the latter as $n$ paths (thus a fuzzer can easily manipulate this metric). In addition, the instability of this metric can be manifested by suppling to AFL all seeds found previously by AFL -- often, the AFL will keep much smaller number of seeds (in some cases only 50\% of the supplied seeds), because the path coverage metric is dependent on the order of processing the seeds.  
Therefore, seed count should not be considered a reliable, standalone metric.

%% file: chapters/eval-2-coverage.tex

\subsection{Coverage}
\label{sec:eval:coverage}

\begin{figure}
    \centering
    \includegraphics[width=0.45\textwidth]{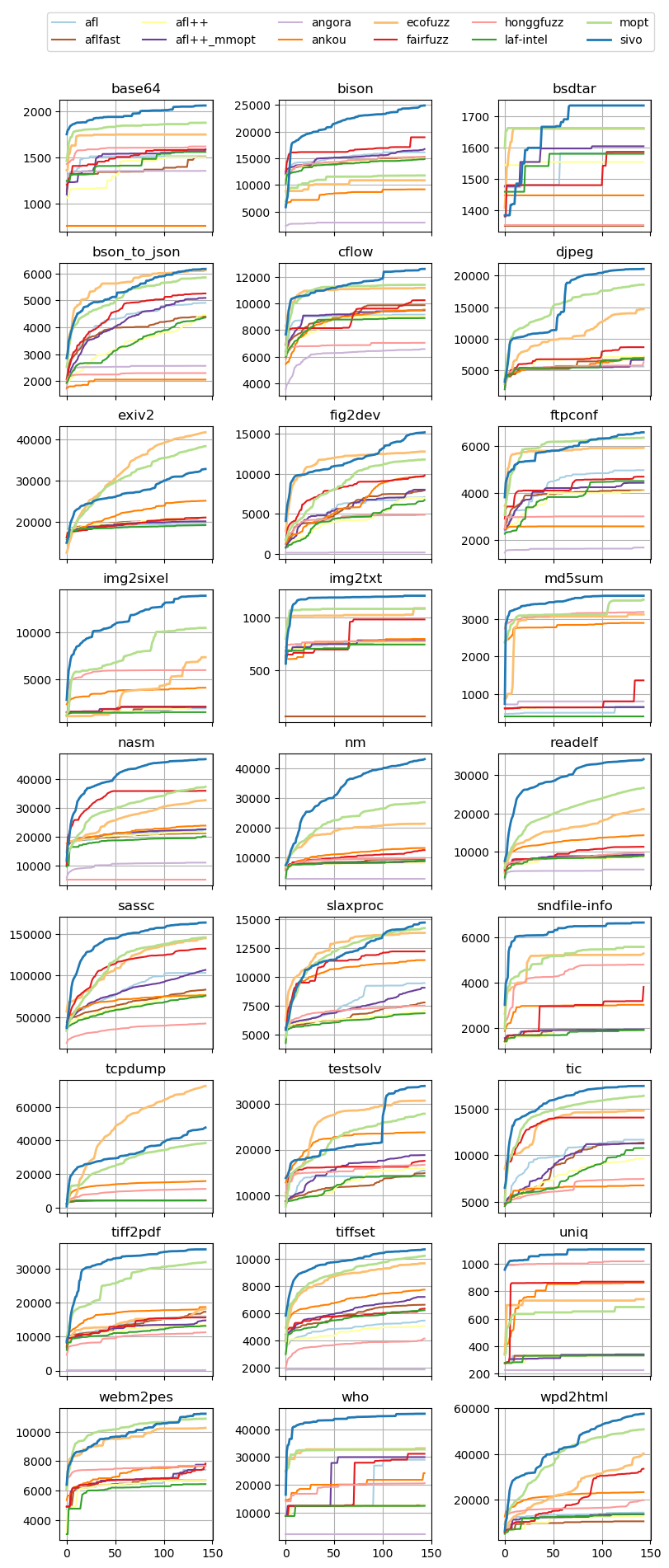} 
    \caption{Coverage for all fuzzers during 12 hours of fuzzing (in 5 min increments).  }
    \label{figure:12hcoverage}
\end{figure}

We run all 12 fuzzers for 12 hours each, and record the coverage discovered during the fuzzing. The results are reported in Figure~\ref{figure:12hcoverage}. We can see that at the end, \fazer provides the best coverage for 25 out of the 27 programs. 
On average \fazer produces $11.8\%$ higher coverage than the next best fuzzer when analyzed individually for each program. In direct comparison to fuzzers, \fazer outperforms the next best fuzzer MOpt by  $20.2\%$, and EcoFuzz by $30.6\%$, and outperforms the baseline AFL by producing  $180\%$ increase in coverage.
For most of the programs, our fuzzer very soon  establishes as the top fuzzer. In fact, the time frame needed to create advantage is so short, that the improved coverage refinement of Section~\ref{sec:refcoverage} has still not kicked in, whereas the MAB optimization of Section~\ref{sec:learning} had barely any time to feed enough data back to the MABs. Thus, arguably the early advantage of \fazer is achieved due to the parametrize paradigm, as well as the remaining two refinements (\taint and the system solver method).

\begin{figure}
    \centering
    \includegraphics[width=0.45\textwidth]{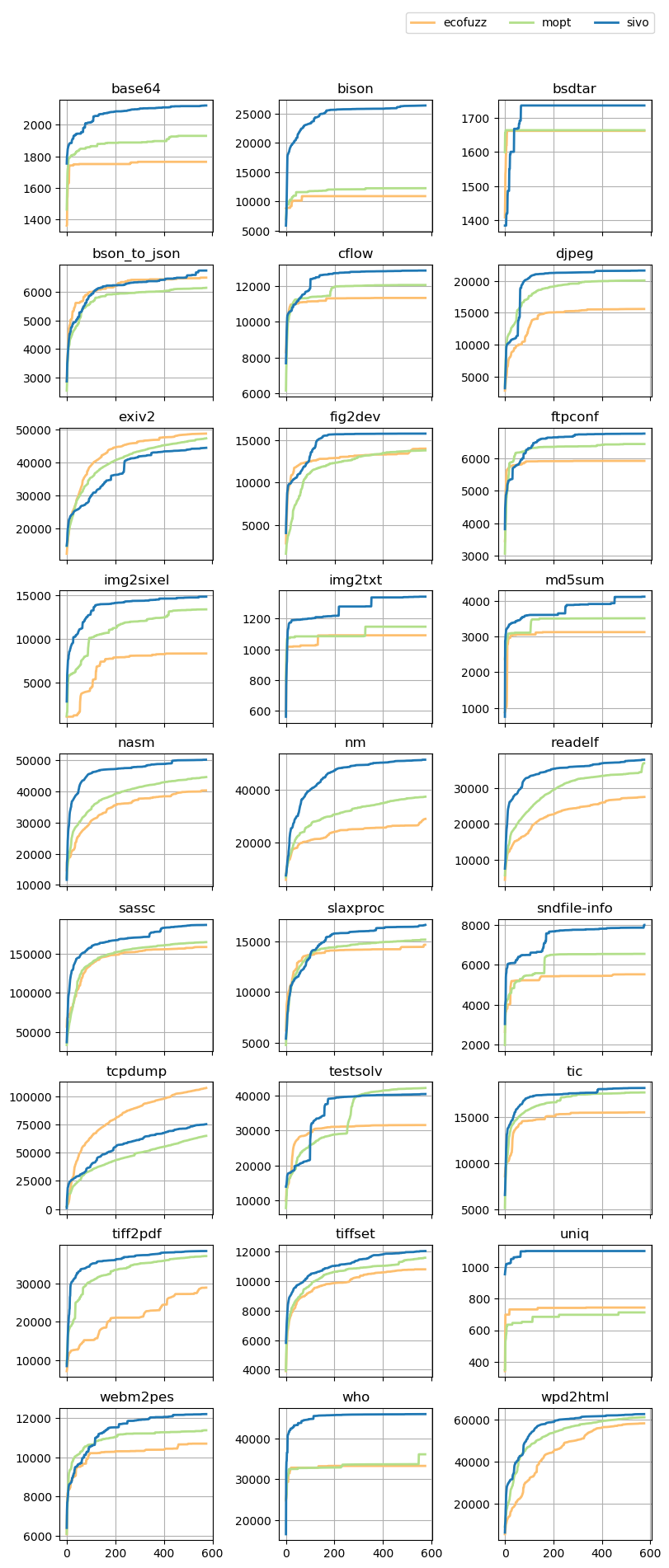}
    \caption{Coverage for top three fuzzers \fazer, MOpt, EcoFuzz during 48 hours of fuzzing.}
    \label{figure:48hcoverage}
\end{figure}

We test the top three fuzzers \fazer, MOpt, and EcoFuzz on 48-hour runs and report the obtained coverage  in Figure~\ref{figure:48hcoverage}.  
We see that \fazer is the top fuzzer for 24 of the programs, with $13.4\%$ coverage increase on average with respect to the next best fuzzer for each program, and $15.7\%$, and $28.1\%$ with respect to MOpt and EcoFuzz.
In comparison to the 12-hour runs, the other two fuzzers managed to reduce slightly the coverage gap, but this is expected (given sufficient time all fuzzers will converge).  
However, the gap is still significant and \fazer provides consistently better coverage. 

%% file: chapters/eval-3-vulnerabilities.tex

\subsection{Vulnerabilities}
\label{sec:eval:vulnerabilities}
\input{chapters/eval-3-vul-table}

We summarize the number of vulnerabilities found by each fuzzer on $25$ programs during the $12$-hour runs in Table~\ref{tbl:vul}. (We removed two programs from Table~\ref{tbl:vul}, as none of the fuzzers finds vulnerabilities for them.)
Out of $25$ evaluated programs, \fazer is able to find the maximal number of vulnerabilities in $18$ programs  ($72$\%).
For comparison, the next best fuzzer MOpt holds top positions in $11$ programs ($44$\%) in terms of vulnerability discovery. This indicates that \fazer is significantly more efficient at finding vulnerabilities than the remaining candidate fuzzers as well.
However, \fazer achieves less top positions in discovery of vulnerabilities compared to code coverage, but this is not unusual as the objective of \fazer is code coverage, and the correlation between produced coverage and found vulnerabilities is not necessarily strong~\cite{klees2018evaluating,inozemtseva2014coverage}. 

We also measure and report in Table~\ref{tbl:vul} the number of vulnerabilities unique to each fuzzer, i.e. bugs that are found only by one fuzzer, and not by any other. 
This metric signals distinctiveness of each fuzzer---the greater the number of unique vulnerabilities, the more distinct the fuzzer is on vulnerability detection. 
Out of $25$ programs, \fazer discovers at least one unique vulnerability in $11$ programs.
In total, \fazer finds $31$ unique vulnerabilities, while the next best fuzzer is Honggfuzz~\cite{honggfuzz} with $21$ vulnerabilities.

%% file: chapters/eval-3-vul-table.tex

\begin{table*}[!ht]
\centering
\caption{ The number of found vulnerabilities. The number of unique vulnerabilities (when non-zero) are reported after "/". 
"-'' indicates failure to instrument/run the program. "\#Vuln." , "\#Vuln. uniq" give the number of all and the number of unique vulnerabilities, respectively. 
"\#Top vuln." shows the number of programs for which the fuzzer finds the maximal number of vulnerabilities. 
"\#Prog. uniq" shows the number of programs for which the fuzzer finds some unique vulnerability.  
} 
\begin{adjustbox}{width=1\textwidth}
\begin{tabular}{l|c|c|c|c|c|c|c|c|c|c|c|c}
\hline
\multirow{2}{*}{Application} & \multicolumn{12}{c|}{Fuzzer} \\ \cline{2-13} 
 & AFL & AFL++ & AFL++\_mopt & AFLFast & FairFuzz & LAF-Intel & MOpt & EcoFuzz & Honggfuzz & Angora & Ankou & \fazer \\ \hline \hline
base64 & 2 & 2 & 2 & 2 & 2 & 2 & 2 & 2 & 2 & 2 & 1 & 2 \\ 
bison & 3 & 3 & 3 & 3 & 4/1 & 3 & 4/1 & 2 & 2 & 1 & 3 & 2 \\ 
bson\_to\_json & 2 & 1 & 1 & 1 & 2 & 2 & 2 & 1 & 1 & 2 & 1 & 2 \\ 
cflow & 2 & 1 & 1 & 2 & 2 & 1 & 5 & 3 & 2 & 1 & 3 & 6/1 \\ 
exiv2 & 6 & 5 & 6 & 5 & 6 & 6 & 11/3 & 0 & - & - & 8 & 8 \\ 
fig2dev & 29/1 & 24 & 29 & 26 & 30/1 & 22 & 35 & 30/2 & 43/4 & 1 & 40 & 59/7 \\ 
ftpconf & 2 & 2 & 2 & 2 & 2 & 2 & 2 & 2 & 2 & 2 & 2 & 2 \\ 
img2sixel & 1 & 1 & 1 & 1 & 1 & 0 & 16/1 & 12/1 & 15/3 & - & 7 & 22/6 \\ 
img2txt & 2 & 2 & 2 & 0 & 4 & 2 & 8/2 & 5/1 & 3 & - & 7/3 & 10/5 \\ 
md5sum & 1 & 1 & 1 & 1 & 2/1 & 1 & 1 & 1 & 1 & 1 & 1 & 1 \\ 
nasm & 4 & 4 & 5 & 4 & 8 & 4 & 10 & 8 & 2 & 5/1 & 9 & 13/1 \\ 
nm & 4 & 3 & 3 & 4 & 4 & 4 & 6/1 & 5 & 3 & 0 & 4 & 6/1 \\ 
readelf & 1 & 1 & 1 & 1 & 1 & 1 & 1 & 1 & 1 & 2/1 & 1 & 1 \\ 
sassc & 1 & 1 & 1 & 1 & 2 & 1 & 2 & 2 & 1 & - & 1 & 5/3 \\ 
slaxproc & 4 & 3 & 3 & 3 & 3 & 3 & 4 & 3 & 3 & - & 6/2 & 5/1 \\ 
sndfile-info & 0 & 0 & 0 & 0 & 3 & 0 & 8/2 & 6 & 13/6 & - & 1 & 7 \\ 
tcpdump & 0 & 0 & 0 & 0 & 0 & 0 & 3 & 1 & 1 & - & 1 & 7/3 \\ 
testsolv & 6 & 6 & 6 & 6 & 6 & 6 & 7 & 8/1 & 14/8 & - & 6 & 9/2 \\ 
tic & 2 & 1 & 2 & 1 & 2 & 2 & 3 & 2 & 2 & - & 0 & 3 \\ 
tiff2pdf & 2 & 2 & 1 & 2 & 1 & 2 & 4 & 3 & 1 & 0 & 3 & 4 \\ 
tiffset & 1 & 1 & 1 & 1 & 1 & 1 & 1 & 1 & 1 & 0 & 1 & 1 \\ 
uniq & 1 & 1 & 1 & 1 & 1 & 1 & 2 & 3 & 7 & 1 & 2 & 7 \\ 
webm2pes & 1 & 1 & 1 & 1 & 1 & 1 & 2 & 2 & 1 & - & 1 & 3/1 \\ 
who & 1 & 1 & 1 & 1 & 1 & 1 & 7 & 3 & 6 & 0 & 3 & 7 \\ 
wpd2html & 0 & 0 & 0 & 0 & 0 & 0 & 1/1 & 1 & 0 & - & 1 & 1 \\ \hline \hline
\#Vuln.        & 78 & 67 & 74 & 69 & 89 & 68 & 147 & 107 & 127 & 18 & 113 & 193 \\ 
\#Top vuln. & 4 & 3 & 3 & 3 & 6 & 4 & 11 & 4 & 6 & 4 & 4 & 18 \\ \hline
\hline
\#Vuln. uniq& 1 & 0 & 0 & 0 & 3 & 0 & 11 & 5 & 21 & 2 & 5 & 31 \\ 
\#Prog. uniq& 1 & 0 & 0 & 0 & 3 & 0 &  7 & 4 &    4 & 2 & 2 & 11 \\ \hline

\end{tabular}
\end{adjustbox}
\label{tbl:vul}
\end{table*}

%% file: chapters/eval-4-stats.tex

\subsection{Performance of Refinements}
\label{sec:eval:stats}

We evaluate the four refinements individually, in terms of their impact and necessity.  
To assess the impact of a refinement, i.e. to estimate how much it helps to advance the fuzzer,  we compare the performance of the baseline version of \fazer (where all four refinements have been removed) to the baseline version with the one refinement added on.  
On the other hand, to assess the necessity of a refinement, i.e. to estimate how irreplaceable in comparison to the other three refinements it is, we compare the full version of \fazer to the version with a single refinement removed. 
We note that all refinements aside for the \textsf{Parametrize-Optimize} strategy, can be assessed reasonably well because it is easy to switch them on or off in the fuzzer. The same holds for \textsf{Optimize}, but not for \textsf{Parametrize}.  As \fazer is built from scratch with many new fuzzing subroutines that are not necessarily present in AFL, it is not clear which fuzzing subroutines and which of their variations need to be removed in the baseline. Therefore, we only assess \textsf{Optimize}, and consider \textsf{Parametrize} to be part of the baseline.  
%
We fuzz the 25 programs (on which \fazer outperformed all other 11 fuzzers) for 12 hours, and compare the found coverage to the coverage produced by the complete version of \fazer. In Table~\ref{tbl:benefits}, we provide the comparisons (as a percentage drop of the coverage) of the versions. 
We also give the data about the performance of the best non-\fazer fuzzer for each program (see the column \textsf{Best NoneSivo}).  In the last row of the table we summarize the number of programs on which the considered version of the fuzzer is able to out-perform all of the remaining 11 none-\fazer fuzzers (for reference, for \fazer this number is 25).  
%
\input{chapters/eval-4-stats-table}

A few observations are evident from the Table~\ref{tbl:benefits}: 
%

\begin{itemize}
\item \textbf{\textsf{Parametrize alone is valuable}}.
The baseline \textsf{SivoBase}, i.e. the version of the fuzzer that does not have any of the four refinements aside from Parametrize, already performs well. It is able to achieve the most coverage for 9 of the 25 considered programs. Hence, just by introducing new fuzzing subroutines and their variations, the fuzzer is able to outperform in terms of coverage the other 11 fuzzers on  36\% of the fuzzed programs.    
%
\item \textbf{\textsf{Optimize} has a strong impact}.
Among the four refinements, Optimize  has the strongest impact. It helps the baseline fuzzer to add 10 top stops resulting in 19 top positions (refer to \textsf{SivoBase+Opt} column in Table~\ref{tbl:benefits}), thus leading to most coverage in comparison to the other 11 none-\fazer fuzzers on 76\% of the programs.  On the other hand, \fazer without Optimize (refer to \textsf{\fazer-Opt}), loses 14 top positions, i.e. the fuzzer loses the top spot for 56\% of the  programs. Moreover, this refinement effects all of the fuzzed programs, with the exception of a few. The effect is significant---the coverage drop when this refinement is not present is at least 10\% and sometimes more than 30\%. 

\item \textbf{\textsf{\taint} has a moderate to low impact}.
This refinement,  denoted as \textsf{FI} in the Table~\ref{tbl:benefits}, helps the baseline fuzzer to add two top spots. On the other hand, \fazer without \taint, i.e. with only the FTI engine present, loses three top spots. \taint has a strong variance  (refer to the \textsf{Sivo-FI} column) in terms of providing additional coverage and most fuzzed programs either benefit largely, or have no benefit at all. This is not unexpected, because the true benefit of \taint is manifested in programs that accept large inputs and that have branches that depend on all of those inputs.  
\item \textbf{\textsf{Solving systems of interval} (SI) has a strong to moderate impact}. It adds 4 top stops to the baseline, and removes 7 top spots from the complete version of \fazer. It provides consistent benefits to the fuzzer -- for most of the fuzzed programs \textsf{SI} produces extra coverage. Presumably, this is based on the fact that most programs do have branches based on integer inequalities and that use direct copy of input bytes.

\item \textbf{\textsf{Accurate coverage} (AC) has a moderate to low impact}. 
This refinement does not have a strong impact on providing top positions (no jumps after adding it to the baseline, and lost 4 positions when removing it from \fazer), but it  gives well balanced improvements in coverage to the fuzzer. 
\end{itemize}

%% file: chapters/eval-4-stats-table.tex

\begin{table*}[!ht]
  \centering
    \caption{ Percentage drop in coverage of fuzzers in comparison to \fazer.  When no drop occurs, the cells are empty.}
\sffamily
\setlength{\tabcolsep}{5pt}
%

\begin{tabular}{l||r|r|r|r|r|r|r|r|r|r|r}
\hline
\multirow{2}{*}{Application }   & \multicolumn{10}{c}{Fuzzer} \\ \cline{2-11}  
& \rotatebox[origin=l]{90}{Best NoneSivo} & \rotatebox[origin=l]{90}{SivoBase} & \rotatebox[origin=l]{90}{SivoBase+Opt} & \rotatebox[origin=l]{90}{SivoBase+FI} & \rotatebox[origin=l]{90}{SivoBase+SI} & \rotatebox[origin=l]{90}{SivoBase+AC} & \rotatebox[origin=l]{90}{Sivo-Opt} & \rotatebox[origin=l]{90}{Sivo-FI} & \rotatebox[origin=l]{90}{Sivo-SI} & \rotatebox[origin=l]{90}{Sivo-AC} \\
\hline
 \textsf{base64} & 9.1 & 7.2 & 2.4 & 7.2 & 4.7 & 7.2 & 3.3 & 3.3 & 8.4 & 1.7 \\
 \textsf{bison} & 23.9 & 23.1 & & 23.1 & 23.1 & 23.1 & 33.4 & & & 4.9 \\
 \textsf{bsdtar} & 4.1 & & & & & & 0.4 & 0.4 & 0.4 & \\
 \textsf{bson\_to\_json} & 0.8 & 18.9 & 3.0 & 18.9 & 16.1 & 18.9 & 17.1 & & 4.6 & 2.9 \\
 \textsf{cflow} & 9.5 & 10.9 & 4.6 & 10.9 & 10.9 & 10.9 & 13.4 & & 3.8 & 5.0 \\
 \textsf{djpeg} & 11.9 & 23.9 & 23.9 & 23.9 & 23.9 & 21.2 & 33.7 & & 15.4 & 22.0 \\
 \textsf{fig2dev} & 15.9 & 13.3 & & 13.3 & 13.3 & 13.3 & 23.0 & & 1.8 & 6.1 \\
 \textsf{ftpconf} & 3.5 & 10.5 & 0.6 & 9.8 & 9.9 & 10.5 & 12.3 & & & 1.4 \\
 \textsf{img2sixel} & 24.7 & 21.9 & 8.0 & 21.5 & 15.9 & 21.9 & 19.9 & 3.8 & 7.7 & 0.6 \\
 \textsf{img2txt} & 9.8 & 9.3 & 9.3 & 9.3 & 9.3 & 8.9 & 8.6 & 8.9 & 7.9 & 10.3 \\
 \textsf{md5sum} & 2.9 & 14.8 & 14.2 & 14.8 & 0.6 & 6.4 & 4.6 & & 12.3 & \\
 \textsf{nasm} & 20.3 & 27.7 & 0.5 & 27.7 & 27.0 & 27.7 & 39.8 & & 3.7 & 0.6 \\
 \textsf{nm} & 33.6 & 11.8 & 11.8 & 11.8 & 6.6 & 11.8 & 15.9 & 43.6 & 27.4 & 18.8 \\
 \textsf{readelf} & 22.0 & 15.2 & & 7.7 & 5.0 & 7.1 & 7.8 & 1.9 & 0.1 & \\
 \textsf{sassc} & 10.9 & 25.4 & & 25.4 & 21.5 & 23.5 & 34.6 & & & \\
 \textsf{slaxproc} & 3.3 & 34.9 & & 31.1 & 28.0 & 30.3 & 38.5 & & 1.9 & \\
 \textsf{sndfile-info} & 16.2 & 17.2 & 10.8 & 17.2 & 11.7 & 17.2 & 6.6 & & 18.6 & 1.7 \\
 \textsf{testsolv} & 9.4 & 43.0 & 33.1 & 42.3 & 10.9 & 24.6 & 33.3 & 34.3 & 37.2 & 10.6 \\
 \textsf{tic} & 6.0 & 16.9 & & 16.9 & 16.7 & 13.7 & 19.7 & & & 0.1 \\
 \textsf{tiff2pdf} & 10.5 & 2.4 & 2.4 & 2.4 & 2.0 & 0.3 & 3.7 & & & \\
 \textsf{tiffset} & 4.4 & 8.9 & 7.8 & & 8.9 & 8.9 & & 0.3 & & \\
 \textsf{uniq} & 7.8 & 16.4 & 0.4 & 16.4 & 3.1 & 16.4 & & 0.2 & 4.8 & \\
 \textsf{webm2pes} & 3.0 & 14.0 & & 12.6 & 14.0 & 14.0 & 12.2 & 7.1 & 6.6 & \\
 \textsf{who} & 27.3 & 29.3 & 13.6 & 23.3 & 17.4 & 27.6 & 2.7 & 9.5 & 35.9 & 10.5 \\
 \textsf{wpd2html} & 11.8 & 27.1 & 13.6 & 27.1 & 27.1 & 27.1 & 48.9 & 6.4 & 0.3 & 5.2 \\
\hline\hline
Top positions  &  & 9  & 19  & 11  & 13  & 9  & 11  & 22  & 18  & 21 \\
\end{tabular}

  \label{tbl:benefits}
\end{table*}

%% file: chapters/eval-5-explain.tex

\subsection{The Cause Of Observed Benefits}
\label{sec:eval:explain}

It is important to understand and explain why certain fuzzing techniques (or in our case refinements) work well. In Section~\ref{sec:eval:stats} we speculate about the type of programs that can be fuzzed well with some of the refinements. Showing this conclusively, however, is difficult. 
Table~\ref{tbl:benefits} shows the percentage drop in coverage observe, per application, obtained by adding and removing one-by-one each of our proposed refinements. 
%
However, attributing the cause of improved performance to individual refinements based on such coarse empirical data could be misleading. This is because we are measuring the joint outcome of mutually-dependent fuzzing strategies. We cannot single out the cause of an observed outcome and attribute it to each strategy, since the strategies mutate the internal state that others use.
We thus only coarsely estimate their impact via our empirical findings and speculate that these results extend to other programs.

%% file: chapters/related-work.tex

\section{Related Work}
\label{sec:related}
Grey-box fuzzers, starting from the baseline AFL~\cite{afl}, have been the backbone of modern, large-scale testing efforts. 
The AFL-family of fuzzers (e.g. AFLGo~\cite{bohme2017directed}, AFLFast~\cite{bohme2017coverage}, LAF-Intel~\cite{lafintel}, MOpt~\cite{lyu2019mopt}, and MTFuzz~\cite{she2020mtfuzz}) improve upon different aspects of the baseline fuzzer. 
For instance, instead of randomly selecting mutation strategy, MOpt~\cite{lyu2019mopt} uses particle swarm optimization to guide the selection. MTFuzz~\cite{she2020mtfuzz} trains a multiple-task neural network to infer the relationship between program inputs and different kinds of edge coverage to guide input mutation.
Similarly, for the seed selection, AFLFast~\cite{bohme2017coverage} prioritizes seeds that exercise low-probability paths, CollAFL~\cite{gan2018collafl} prioritizes seeds that have a lot of not-yet inverted branches, and EcoFuzz~\cite{yue2020ecofuzz} uses multi-armed bandits to guide the seed selection. 
Common feature for all current fuzzers from the AFL-family is that they optimize at most one of the fuzzing subroutine\footnote{This refers to optimization only -- some fuzzers improve (but not optimize) multiple fuzzing subroutines.}. 
In contrast, \fazer first parameterizes all aspects, i.e. introduces many variations of the fuzzing subroutines, and then tries to optimize all the selection of parameters. Even the seed selection subroutines of EcoFuzz and \fazer  differ, despite both using multi-armed bandits: EcoFuzz utilizes MAB to select candidate seed from the pool, whereas \fazer uses MAB to decide on the selection criterion and the pool of seeds.

Several grey-box fuzzers deploy data-flow fuzzing, i.e. infer dependency of branches on input bytes and use it to accomplish more targeted branch inversion. 
VUzzer~\cite{rawat2017vuzzer}, Angora~\cite{chen2018angora}, BuzzFuzz~\cite{ganesh2009taint} and Matryoshka~\cite{chen2019matryoshka} use a classical dynamic taint inference engine (i.e. track taint propagation) to infer dependencies. Fairfuzz~\cite{lemieux2018fairfuzz}, ProFuzzer~\cite{you2019profuzzer}, and Eclipser~\cite{choi2019grey} use lighter engine and infer partial dependency by monitoring the execution traces of the seeds. RedQueen~\cite{aschermann2019redqueen} and Steelix~\cite{li2017steelix} can infer only dependencies based on exact (often called direct)  copies of input bytes in the branches, by mutating individual bytes.  
Among grey boxes, the best inference in terms of speed, type, and accuracy is achieved by GreyOne~\cite{gan2020greyone}. Its engine called FTI is based on mutation of individual bytes (thus fast because it does not track taint propagation) and can detect dependencies of any type (not only direct copies of input bytes). 
FTI mutates bytes one by one and checks on changes in variables involved in branch conditions (thus accurate because it does not need for the whole branch to flip, only some of its variables). 
\fazer inference engine \taint improves upon FTI and provides exponential decrease in the number of executions required to infer the full dependency, at a possible expense of accuracy. Instead of testing bytes one by one, \taint uses probabilistic group testing and reduces the number of executions.

Data-flow grey boxes accomplish targeted branch inversion by randomly mutating the dependent bytes. 
A few fuzzers deploy more advanced strategies: Angora~\cite{chen2018angora} uses gradient-descent based mutation, Eclipser~\cite{choi2019grey} can invert efficiently branches that are linear or monotonic, and GreyOne~\cite{gan2020greyone} inverts branches by gradually reducing the distance between the actual and expected value in the branch condition. 
Some fuzzers, such as RedQueen and Steelix invert branches by solving directly the branch conditions based on equality (called magic bytes). \fazer can solve more complex branch inversion conditions that involve inequalities, without the use of SAT/SMT solvers. On the other hand, white boxes such as KLEE~\cite{cadar2008klee}, and hybrid fuzzers such as Driller~\cite{stephens2016driller} and QSYM~\cite{yun2018qsym}, use symbolic execution that relies on SMT solvers (thus it may be  slow) to perform inversions in even more complex branches. 
The hybrid fuzzer Pangolin~\cite{huang2020pangolin} uses linear approximations of branch constraints (thus more general than our intervals) called polyhedral path abstraction and later it utilizes them to efficiently sample solutions that satisfy path constraints. To infer the (more universal) linear approximations, Pangolin uses a method based on SMT solver. On the other hand, \fazer infers the (less universal) intervals with a simpler method. 

The AFL-family of fuzzers as well as many other grey boxes track \emph{edge coverage}. In addition, the AFL-family uses bucketization, i.e. besides edges, they track the counts of edges and group them in buckets that have ranges of powers of two. For practical purposes AFL does not record the precise edges (this will require storing whole execution traces which may be slow), but rather it works with hashes of edges (which is quite fast). The process of hashing may introduce collisions as noted by CollAFL~\cite{gan2018collafl}. 
To avoid such collisions, CollAFL proposes during compilation to choose the free parameters of the hashing function non-randomly, and according to a specific strategy. AFL++~\cite{fioraldi2020afl++} uses a similar idea and provides an open-source implementation based on link-time instrumentation.  
In addition, AFL++,LibFuzzer~\cite{serebryany2016continuous}, and Honggfuzz~\cite{honggfuzz} use so-called sanitizer coverage available in LLVM starting from version 11 to prevent collisions by assigning the free parameters during runtime. 
On the other hand, \fazer solution is to switch between different hashing functions during the fuzzing (i.e. at runtime). 
Instead of tracking edge coverage, a few fuzzers such as Honggfuzz~\cite{honggfuzz}, VUzzer~\cite{rawat2017vuzzer} and LibFuzzer~\cite{serebryany2016continuous}  track block coverage.  
Moreover, the grey-box fuzzer TortoiseFuzz~\cite{wang2020not} uses alternative coverage measurement metric (assigns different weights to edges based on their potential security impact) to prioritize testcases, and achieves higher rate of vulnerability detection.

%% file: chapters/conclusion.tex

\section{Conclusion}
\label{sec:conclusion}

We have presented four refinements for grey boxes that boost different fuzzing stages. 
First, we have shown fast taint engine that requires only logarithmic number of tests in the number of input bytes to infer the dependencies of branches on inputs. Second, we have provided an efficient method for inverting branches when they depend trivially on input bytes and their conditions are based on integer inequalities. 
Third, we have proposed an improved coverage tracking methods that are easy to implement. 
Finally, we have show the parametrize-optimize paradigm that allows fuzzers to be more flexible in adopting to target programs and thus to effectively increase coverage.   
We have implemented the refinements in a fuzzer called \fazer. 
In comparison to 11 other popular grey-box fuzzers, \fazer scores highest with regards to coverage and to number of found vulnerabilities.

%% file: chapters/ack.tex
\section{Acknowledgments}
We thank our shepherd Erik van der Kouwe for his helpful feedback. Abhik Roychoudhury, Zhijingcheng Yu, Shin Hwei Tan, Lu Yan, Andrea Fioraldi,  and the anonymous reviewers gave us valuable comments and improvements on this work, for which we are thankful.
All opinions expressed in this paper are solely those of the authors. This research is supported in part by the Crystal Centre at NUS and by the research grant DSOCL17019 from DSO in Singapore.

%% file: paper-appendix.tex

\input{chapters/appendix-coverage-combinatorial}

\input{chapters/appendix-code}

\input{chapters/appendix-instrumentation-overhead}

\FloatBarrier
\input{chapters/appendix-programs}

\input{chapters/appendix-coverage-plots}

%% file: chapters/appendix-coverage-combinatorial.tex

\section{Analysis of Multi-Label Block Assignment}
\label{sect:appendix:multibloc}

When each basic block $B_i$ is assigned an $n$-bit random label $L_i$ , by the birthday paradox collisions on hashed edges $L_j \lll \oplus L_k$  will appear once the number of edges $2^t$ reaches around $2^t \ge 2^\frac{n}{2}$. Roughly, the expected number of collisions is around $2^t \cdot 2^ t / 2 ^ n = 2 ^ {2t - n} $. If we assign an additional (second) label to each basic block, then there will be roughly $2 ^ {2t - n} $ colliding edge hashes on the second label as well. However, among the colliding hash edges on the first and the second label, there will be only $2 ^{2t -n } \cdot 2 ^ {2t -n } / 2 ^ t = 2 ^ { 3t - 2n }$ common edges. Thus, if $3t - 2n < 0$, i.e. $ t < \frac {2n}{3}$, then on average there will not be any common edges, and hence for at least one of the assignments (either the first or the second), each edge will have unique hash. Similar analysis applies for larger number of labels. In general, if each block is assigned $m$ labels of $n$ random bits each, and the number of edges is smaller than $2^{\frac{m}{m+1}n}$, then on average each hashed edge will be unique at least for one of the labels. 

We can also obtain a strict analysis is of the collision-free multi-label assignment.  To do so, we compute the probability that each additional hashed edge does not collide with any other previous hashed edge at least on one of the labels. If there are already $k$ such hashed edges, then the probability that the next $k+1$ will also be good is $1 - (\frac{k}{2^n})^m$.  This is computed from the opposite event (that the $k+1$-th hash collides  on all $m$ labels with some of the $k$ hashes). 
Thus the probability that all $2^t$ edges will be unique on at least one of the $m$ labels is:
\begin{equation}
\label{eq:labels}
\prod_{k=1}^{2^t-1} \left( 1 - \left(\frac{k}{2^n}\right)^m \right)
\end{equation}
We could not reduce the above formula~(\ref{eq:labels}) further (unlike the case of $m=1$ where Sterling approximation can be used), however we can calculate numerical values of the probability for different values of the parameters $n,t,m$. In the case of \fazer, $n = 16, m=4$, and therefore, even when there are $t=8000$ edges, the probability that all of them will be unique on some label is around 0.701. 
For larger range of computed probabilities refer to Table~\ref{tbl:prob}.

\begin{table}[!ht]
  \centering
  \setlength{\tabcolsep}{4pt}
    \caption{Calculated values of the probability~(\ref{eq:labels})}
  \begin{tabular}{lcccccccc}
 $m/t$ 				& 3000 & 4000 & 5000 & 6000 & 7000 & 8000 & 9000 & 10000  	 	 \\
 \hline 
1  & 0.000  & 0.000  & 0.000  & 0.000  & 0.000  & 0.000  & 0.000  & 0.000  \\ 
2  & 0.123  & 0.007  & 0.000  & 0.000  & 0.000  & 0.000  & 0.000  & 0.000  \\ 
3  & 0.931  & 0.797  & 0.574  & 0.317  & 0.119  & 0.026  & 0.003  & 0.000  \\ 
4  & 0.997  & 0.989  & 0.967  & 0.919  & 0.834  & 0.701  & 0.527  & 0.338  \\ 
5  & 1.000  & 0.999  & 0.998  & 0.994  & 0.984  & 0.965  & 0.929  & 0.871  \\ 
\end{tabular}
  \label{tbl:prob}
\end{table}

%% file: chapters/appendix-code.tex

\section{Pseudo-code of Generic Grey-Box Fuzzer}
\label{sect:appendix:code}

\begin{algorithm}
seed			$\gets$ Sample( ConstantCriterion , Seeds )\;
Taint\_inference(seed)				\tcp*{if dataflow fuzzer}	
\While{time budget left }{
	use\_mut $\gets$ Sample( strategy )\;
	new\_seed $\gets$ Mutate( seed, use\_mut, ConstantParams )\;
	new\_coverage $\gets$ ProduceCoverage(new\_seed)\;
	cov\_increase $\gets$ $\|$ new\_coverage $\setminus$ Coverage $\|$ \;
	\If{ cov\_increase > 0  }{
		Seeds $\gets$ Seeds $\bigcup$ new\_seed \;
		Coverage $\gets$ Coverage $\bigcup$ new\_coverage \;
	}
}
\caption{OneIterationGenericFuzz( Seeds, Coverage )}
\label{alg:iterationGeneric}
\end{algorithm}

\begin{algorithm}
Seeds  		$\gets$ Initial\_seeds\;
Coverage $\gets$ ProduceCoverage(Seeds)\;
\While{ true }{
	OneIterationGenericFuzz( Seeds, Coverage );
}
\caption{GenericFuzzer}
\label{alg:fuzzerGeneric}
\end{algorithm}

%% file: chapters/appendix-instrumentation-overhead.tex

\section{Instrumentation overhead}
\label{sect:appendix:instrumentation}

Each of the 27 evaluated programs is instrumented and compiled with two different passes: lighter and heavier. We measured experimentally the usage and the overhead of the two instrumentations averaged over all 27 tested programs, and  provide the data in Table~\ref{tbl:instrumentation}. 
We give the percentage of usage of the two instrumentations as well as their overhead on top of normal, uninstrumented program and on top of the standard AFL code coverage instrumentation, measured according to the  execution times of the compiled programs. We can see that the lighter instrumentation is 73 \% slower than the uninstrumented program and 39 \% slower than AFL instrumentation, whereas heavier is 270 \%  and 190 \% slower, respectively. On the other hand, the versions with lighter instrumentation are executed far more frequently (78\% vs 22\%), thus we can conclude that on average \fazer introduce 116 \% overhead on top of the uninstrumented programs, and 72 \% on top of AFL.  

\begin{table}[!ht]
  \centering
    \caption{Instrumentation Statistics}
  \begin{tabular}{lrrr}
 Type 				& Usage 	& Overhead  	 	 	& Overhead  	 	 \\
 	 				&  		& uninstr.  			&  AFL  	 \\
 \hline 
 Lighter 				&  $78\%$ 	& $73\%$	& $39\%$	 \\
 Heavier 				&  $22\%$ 	& $270\%$	& $190\%$	 \\
\end{tabular}
  \label{tbl:instrumentation}
\end{table}

%% file: chapters/appendix-programs.tex

\section{Tested Programs}
\label{sect:appendix:programs}

\begin{table}[!ht]
  \centering

  \begin{tabular}{ll>{\footnotesize}l}
  Program & Version & Input line (as in AFL)	\\
 \hline 
 \hline

 \textsf{base64}         & LAVA-M             & -d @@                                       \\
 \textsf{bison}          & bison 3.0.5        & @@                                          \\
 \textsf{bsdtar}         & libarchive 3.4.3   & -acf bsdtar.tar @@                          \\
 \textsf{bson\_to\_json} & libbson 1.8        & @@                                          \\
 \textsf{cflow}          & cflow 1.5          & @@                                          \\
 \textsf{djpeg}          & libjpeg 2.0.90     & -colors 234 -rgb -gif -outfile djp.gif @@ \\
 \textsf{exiv2}          & exiv2 0.27.3       & -pt @@                                      \\
 \textsf{fig2dev}        & fig2dev 3.2.7a     & @@                                          \\
 \textsf{ftpconf}        & libconfuse 3.2.2   & @@                                          \\
 \textsf{img2sixel}      & libsixel 1.8.2     & @@                                          \\
 \textsf{img2txt}        & libcaca 0.99beta19 & @@                                          \\
 \textsf{md5sum}         & LAVA-M             & -c @@                                       \\
 \textsf{nasm}           & nasm 2.14rc15  	  & -f elf64 @@ -o nasm.o                       \\
 \textsf{nm}             & binutils 2.31  	  & -DC @@                                      \\
 \textsf{readelf}        & binutils 2.31  	  & -a @@                                       \\
 \textsf{sassc}          & libsass 3.5        & @@                                          \\
 \textsf{slaxproc}       & libslax 0.22.0     & -c @@                                       \\
 \textsf{sndfile-info}   & libsndfile 1.0.28  & @@                                          \\
 \textsf{tcpdump}        & tcpdump 4.10.0rc1  & -AnetttttvvvXXr @@                          \\
 \textsf{testsolv}       & libsolv 0.7.2      & @@                                          \\
 \textsf{tic}            & ncurses 6.1        & -o tic.out                                  \\
 \textsf{tiff2pdf}       & libtiff 4.0.9      & @@                                          \\
 \textsf{tiffset}        & libtiff 4.0.9      & -s 315 whatever @@                          \\
 \textsf{uniq}           & LAVA-M             & @@                                          \\
 \textsf{webm2pes}       & libwebm 1.0.0.27   & @@ webm2pes.out                             \\
 \textsf{who}            & LAVA-M             & @@                                          \\
 \textsf{wpd2html}       & libwpd 0.10.1      & @@                                          \\

\end{tabular}
  \label{tbl:programs}
\end{table}

%% file: chapters/appendix-coverage-plots.tex

\section{Alternative Coverage}
\label{sect:appendix:coverage}

\begin{figure}
    \centering
    \includegraphics[width=0.45\textwidth]{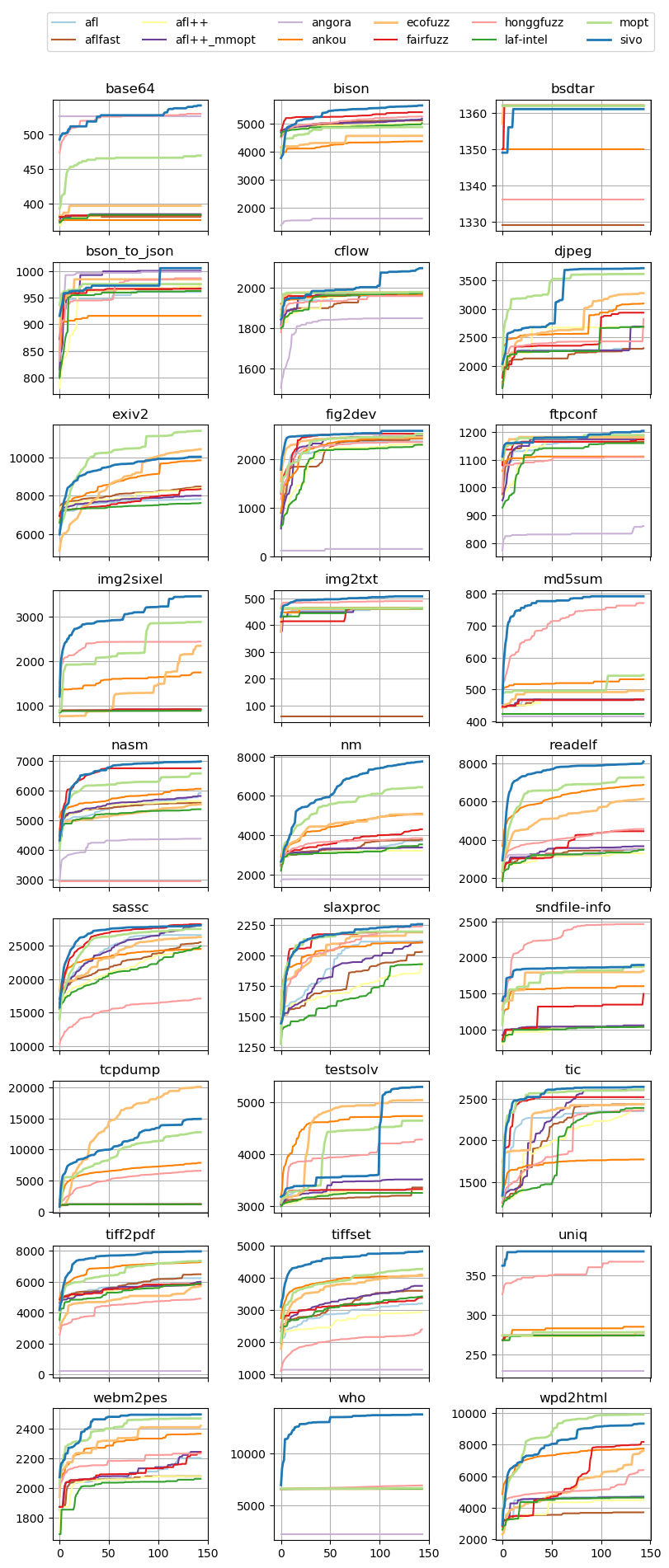} 
    \caption{Edge count for all fuzzers during 12 hours of fuzzing. }
    \label{figure:12hedge}
\end{figure}

\begin{figure}
    \centering
    \includegraphics[width=0.45\textwidth]{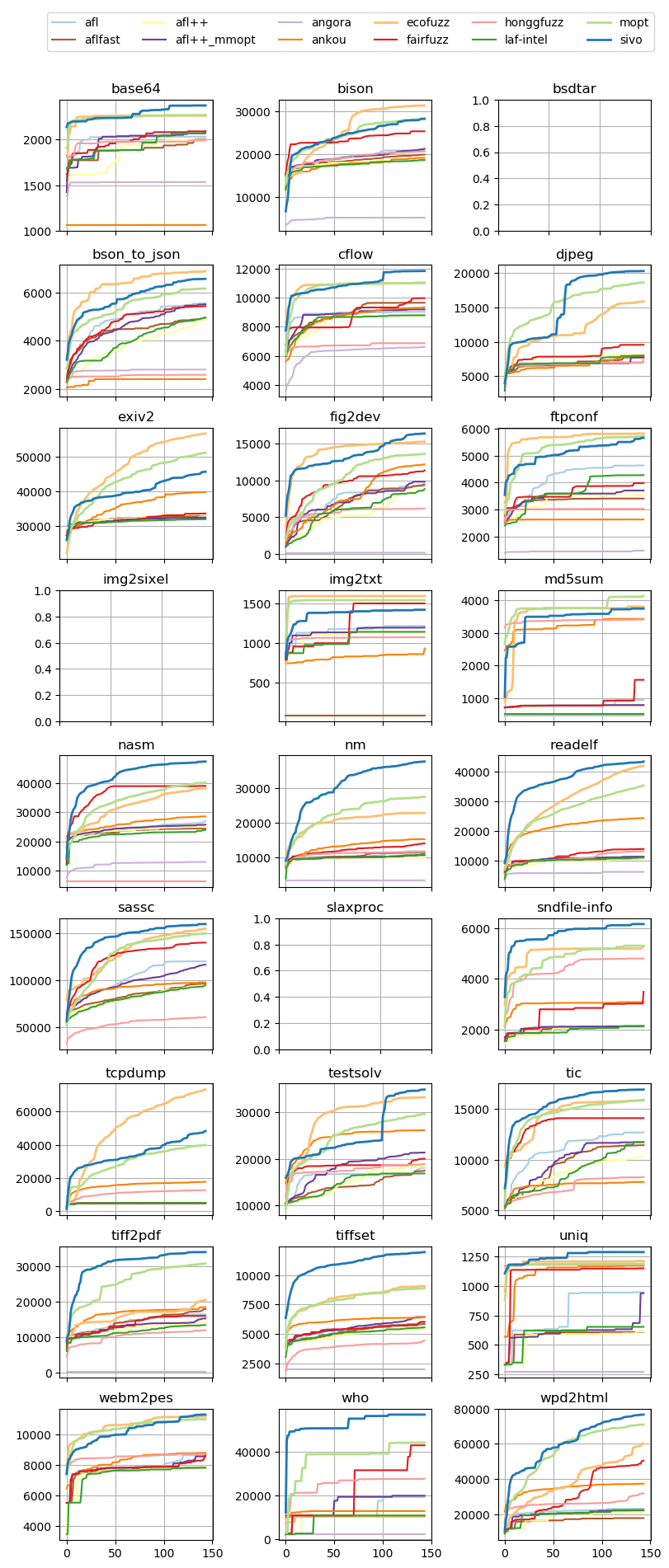}
    \caption{AFL coverage for all fuzzers during 12 hours of fuzzing. }
    \label{figure:12hafl}
\end{figure}

In Figure~\ref{figure:12hedge} we show the simple edge count for fuzzers on 12-hour runs. \fazer is top fuzzer for 21 of the programs in the 12-hour runs. In comparison to Figure~\ref{figure:12hcoverage}, \fazer loses 4 top spots, 2 of which are with a margin of less than 1\%.
In Figure~\ref{figure:12hafl} we show the coverage measured imprecisely as in AFL for fuzzers on 12-hour runs. 
\fazer is top fuzzer for 17 out of 24 programs (we had problems compiling 3 of the programs).  
In comparison to Figure~\ref{figure:12hcoverage}, \fazer loses 6 top spots. This is a result of colliding edge hashes and the inability of AFL coverage engine to detect and handle such collisions.

